\documentclass[cernpreprint,atlasdraft=false,texlive=2017,USenglish,texmf]{atlasdoc}
\usepackage{atlaspackage}
\usepackage{atlasbiblatex}
\usepackage{atlasphysics}
\usepackage[output=false]{atlastodo}
 
\graphicspath{{logos/}{figures/}}
\usepackage{ANA-HDBS-2018-47-PAPER-defs}
\PreprintIdNumber{CERN-EP-2020-067}
\AtlasTitle{Search for Higgs boson decays into two new low-mass
spin-0 particles in the 4$b$ channel with the ATLAS detector using $pp$ collisions at $\sqrt{s}= 13$~TeV}
\AtlasAbstract{
 
This paper describes a search for beyond the Standard Model decays of
the Higgs boson into a pair of new spin-0 particles subsequently
decaying into $b$-quark pairs, $H \rightarrow aa \rightarrow
(b\bar{b})(b\bar{b})$, using proton--proton collision data collected by the
ATLAS detector at the Large Hadron Collider at center-of-mass energy
$\sqrt{s}=13$~TeV. This search focuses on the range $15$~GeV~$\leq m_a \leq
30$~GeV, where the decay products are collimated; it is complementary
to a previous search in the same final state targeting the range
$20$~GeV~$\leq m_a \leq 60$~GeV, where the decay products are well
separated. A novel strategy for the identification of the $a
\rightarrow b\bar{b}$ decays is deployed to enhance the efficiency
for topologies with small separation angles. The search is performed
with 36 fb$^{-1}$ of integrated luminosity collected in 2015 and 2016
and sets upper limits on the production cross-section of $H
\rightarrow aa \rightarrow (b\bar{b})(b\bar{b})$, where the Higgs boson is produced in
association with a $Z$ boson.
}
 
\author{The ATLAS Collaboration}
\AtlasRefCode{HDBS-2018-47}
\AtlasJournalRef{Phys. Rev. D 102 (2020) 112006}
\AtlasDOI{10.1103/PhysRevD.102.112006}
\AtlasJournal{Phys. Rev. D}
\hypersetup{pdftitle={ATLAS document},pdfauthor={The ATLAS Collaboration}}
 
\begin{document}
 
\maketitle
\section{Introduction}
 
The Higgs boson is a particle with a particularly narrow natural width, and its branching fractions to new light particles can be sizable even if they interact weakly with it. Because of this, several new weakly interacting light particles that would not be visible in inclusive searches can be probed by searching for ``beyond the Standard Model'' (BSM) Higgs boson decays at the LHC~\cite{Curtin:2013fra}. These new light particles are predicted in several BSM theories with extended Higgs sectors~\cite{Dobrescu:2000yn,Ellwanger:2003jt,Dermisek:2005ar,Chang:2008cw,Morrissey:2008gm} that address open questions in high-energy physics. Theories with new light particles weakly coupled to the Higgs boson provide an explanation for electroweak baryogenesis~\cite{Profumo:2007wc,Blinov:2015sna} and contain fields that mediate interactions between Standard Model (SM) particles and dark matter~\cite{Silveira:1985rk,Pospelov:2007mp,Draper:2010ew,Ipek:2014gua,Martin:2014sxa}. This paper presents a search for a new spin-0 singlet $a$ that couples to the SM Higgs boson.
 
\begin{figure*}[htbp]
\begin{center}
\includegraphics[width=0.5\textwidth]{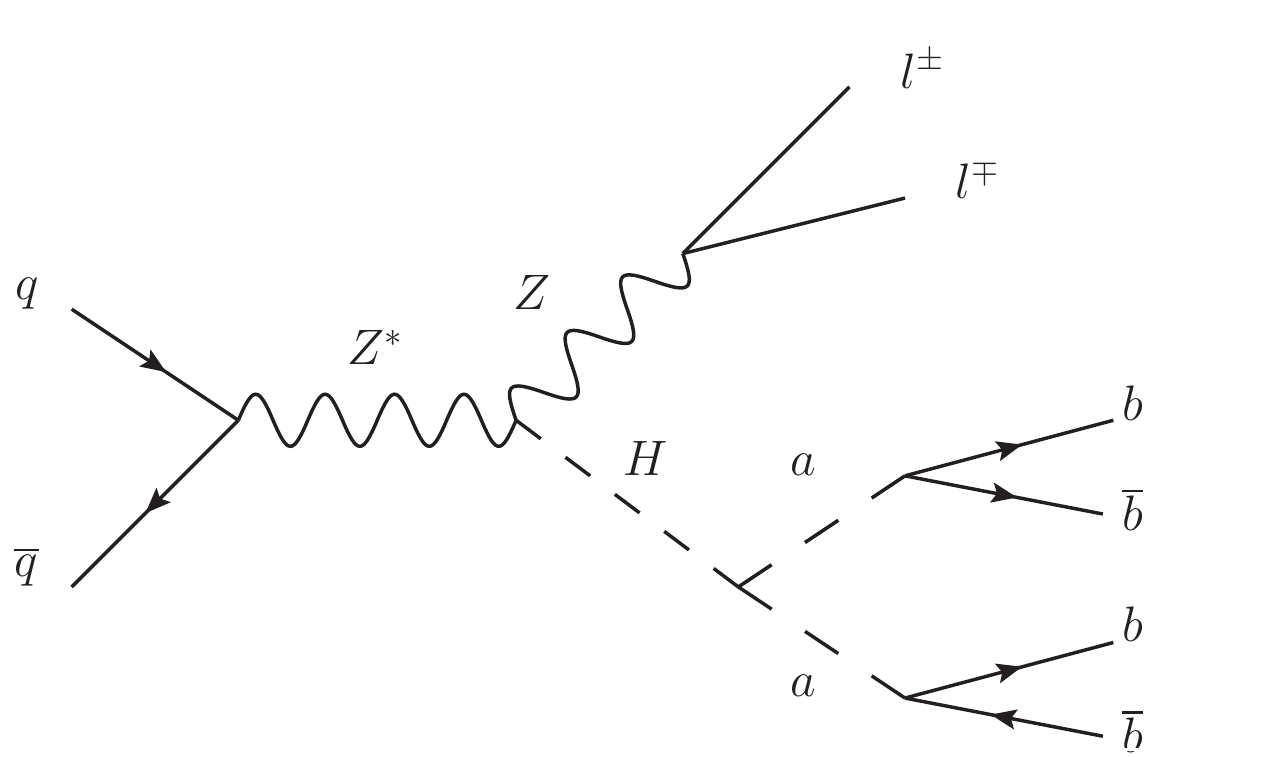}\\
\caption{Representative tree-level Feynman diagram for the $ZH$ production processes with the subsequent decays
$Z\to\ell\ell$ ($\ell=e,\mu$) and $H \to  aa \to (\bbbar)(\bbbar)$.}
\label{fig:feynmandiagrams}
\end{center}
\end{figure*}

When the mass of the spin-0, $m_a$, is less than half of the mass of the Higgs boson, $m_H$, i.e.\ $2m_a < m_H$, the decay $H\to aa$ is kinematically allowed. The search in this paper is performed with events in which each $a$-boson decays into a pair of $b$-quarks, and the Higgs boson is produced in association with a $\Zboson$ boson which decays into electrons or muons. The final state with multiple $b$-quarks has the highest branching ratio in several BSM theories when it is kinematically accessible. The $\Zboson$ boson with leptonic decay provides a simple strategy for triggering and selecting events, as well as powerful background rejection. Figure~\ref{fig:feynmandiagrams} depicts the main production mechanism of the events sought in this paper.
 
The Higgs boson has been observed by the ATLAS and CMS collaborations~\cite{ATLAS_Higgs,CMS_Higgs}. A comprehensive program is being pursued to measure its branching ratios to SM particles and to search for decays into exotic or non-SM particles. Current measurements constrain the non-SM branching ratio of the Higgs boson to be less than approximately 21\% at 95\% confidence level (C.L.) with several assumptions~\cite{Aad:2019mbh}, leaving enough room for exotic Higgs boson decays.
 
ATLAS has previously performed a search where each of the four $b$-quarks was experimentally identified as an individual jet in the detector~\cite{Aaboud:2018iil}. The search set upper limits on the production cross-section of $ZH$, followed by $H\to aa\to (\bbbar)(\bbbar)$, of approximately $0.5$~pb at 95\% C.L.\ for $m_a \gtrsim 30\,\GeV$. However, when the mass of the $a$-boson is small, it is produced with large momentum and the jets created in the hadronization of the two $b$-quarks from a single $\abb$ decay are reconstructed as a single jet in the calorimeter using the standard ATLAS reconstruction algorithms. Because of this, the previous search that covered the range $20\,\GeV\,\leq m_a \leq 60\,\GeV$ rapidly loses efficiency for masses $m_a \lesssim 30\,\GeV$.
 
This article extends the previous analysis in the mass regime $15\,\GeV\,\leq m_a \leq 30\,\GeV$ by relying on a novel strategy for the reconstruction and identification of $\abb$ decays. The article is structured as follows. Section~\ref{sec:detector} describes the relevant features of the ATLAS detector. Section~\ref{sec:datasim} lists the data
collected for this search and details the simulated signal and background event samples that were used to describe the composition of the selected events. Section~\ref{sec:objreco} describes the basic reconstruction and identification of leptons and jets using the ATLAS  detector. Section~\ref{sec:atobb_id} presents the dedicated method for the reconstruction and identification of low-mass $\abb$ decays. Section~\ref{sec:analysis} explains the strategy for event selection and categorization. Section~\ref{sec:systematics} discusses the systematic uncertainties considered in this search, and Section~\ref{sec:result} presents the results. Finally, Section~\ref{sec:conclusion} presents the conclusion.

\section{ATLAS detector}
\label{sec:detector}
 
The ATLAS experiment~\cite{PERF-2007-01,ATLAS-TDR-2010-19,PIX-2018-001} at the LHC is a multipurpose particle detector with a forward--backward symmetric cylindrical geometry and a near \(4\pi\) coverage in
solid angle.\footnote{\AtlasCoordFootnote} It consists of an inner tracking detector surrounded by a thin superconducting solenoid providing a \SI{2}{\tesla} axial magnetic field, electromagnetic and hadron calorimeters, and a muon spectrometer. The inner tracking detector covers the pseudorapidity range \(|\eta| < 2.5\). It consists of silicon pixel, silicon microstrip, and transition radiation tracking detectors.
Lead/liquid-argon (LAr) sampling calorimeters provide electromagnetic (EM) energy measurements with high granularity. A steel/scintillator-tile sampling calorimeter provides hadronic energy measurements in the central pseudorapidity range (\(|\eta| < 1.7\)). The endcap and forward regions are instrumented with LAr calorimeters for EM and hadronic energy measurements up to \(|\eta| = 4.9\). The muon spectrometer surrounds the calorimeters and is based on three large air-core toroidal superconducting magnets with eight coils each. The field integral of the toroids ranges between \num{2.0} and \SI{6.0}{\tesla\metre}
across most of the detector. The muon spectrometer includes a system of precision tracking chambers and fast detectors for triggering. A two-level trigger system is used to select events. The first-level trigger is implemented in custom hardware and uses a subset of the detector information to keep the accepted rate below \SI{100}{\kHz}. This is followed by a software-based trigger that reduces the accepted event rate to \SI{1}{\kHz} on average depending on the data-taking conditions.
 
\section{Dataset and simulated event samples}
\label{sec:datasim}
 
Events are selected from proton--proton ($pp$) collisions collected by the ATLAS detector at the LHC at $\sqrt{s}=13~{\TeV}$ in 2015 and 2016. Only collisions recorded when all relevant subsystems were operational are considered in the analysis. The dataset corresponds to an integrated luminosity of $3.2 \pm 0.1$~fb$^{-1}$ recorded in 2015 and $32.9 \pm 0.7$ fb$^{-1}$ recorded in 2016, for a total of $36.1 \pm 0.8$ fb$^{-1}$~\cite{DAPR-2013-01}. The uncertainty is obtained from the primary luminosity measurements using the LUCID-2 detector~\cite{LUCID2}. The data used for this search were collected using the single-electron or single-muon triggers with transverse momentum ($\pt$) thresholds of 20 (26)~\GeV\ for muons and 24 (26)~\GeV\ for electrons in 2015 (2016)~\cite{TriggerRun2}.
 
Simulated event samples are used to study the characteristics of signal events and to calculate the signal efficiency and acceptance, as well
as for most aspects of the background estimation.  Monte Carlo (MC) samples were produced using the full ATLAS
detector simulation~\cite{Aad:2010ah} based on {\GEANT4}~\cite{Agostinelli:2002hh}. To simulate the effects of simultaneous inelastic collisions (pileup), additional interactions were generated using \textsc{Pythia}~8.186~\cite{Sjostrand:2007gs} with
the A2 set of tuned parameters~\cite{Pythia8Tunes2012} and the MSTW2008LO~\cite{Martin:2009iq} parton distribution function (PDF) set, and overlaid on the simulated hard-scatter event. Simulated events were reweighted to
match the pileup conditions observed in the data. All simulated events are processed through the same reconstruction algorithms and analysis chain as the data. Decays of $b$- and $c$-hadrons were performed by \textsc{EvtGen} v1.2.0~\cite{Lange:2001uf}, except in events simulated with the $\sherpa$ event generator~\cite{Gleisberg:2008ta}.
 
Signal samples of associated Higgs boson production with a $\Zboson$ boson, $pp \to ZH$, were generated with
\POWHEGBOXV{v2}~\cite{Nason:2004rx,Frixione:2007nw,Frixione:2007vw,Alioli:2010xd}
using the CT10 PDF set~\cite{Lai:2010vv} at next-to-leading order (NLO). The sample include gluon-initiated processes at LO. The Higgs boson decay into two spin-0 $a$-bosons and the subsequent decay of each $a$-boson into a pair of $b$-quarks were simulated with \pythia~8.186. The $a$-boson decay was performed in the narrow-width approximation and the coupling to the $b$-quarks is assumed to be that of a pseudoscalar. The information about the parity of the $a$-boson assumed in the simulation is lost in the hadronization of the $b$-quarks and, therefore, the results of this search apply equally to scalars and pseudoscalars. \pythia~8.186 was also used for the parton showering, hadronization, and underlying-event simulation with the A14 tune~\cite{Pythia8Tunes2014}. Signal events were generated for several $a$-boson mass hypothesis: $15\,\GeV$, $17.5\,\GeV$, $20\,\GeV$, $22.5\,\GeV$, $25\,\GeV$, $27.5\,\GeV$, and $30\,\GeV$.
 
The background samples were generated following exactly the same procedure as described in Ref.~\cite{Aaboud:2018iil} and only a summarized description is given here. A sample of top-quark pair events was generated using \POWHEGBOXV{v2}~\cite{Powhegttbar} with the NNPDF3.0NLO PDF set. The parton showers and hadronization were modeled by  \pythia~8.210~\cite{Sjostrand:2014zea} with the A14 tune. To model the \ttbb\ background with better precision, the relative contributions of the different heavy-flavor categories in the $\ttbar$ sample are scaled to match the predictions of an NLO \ttbb\ sample including parton showering and hadronization~\cite{Cascioli:2013era}, generated with \textsc{Sherpa+OpenLoops}~\cite{Gleisberg:2008ta, Cascioli:2011va}, using the procedure described in Ref.~\cite{HIGG-2013-27}.
 
The production of $\Zboson$ bosons in association with jets was simulated with \SHERPA~2.2.1~\cite{Gleisberg:2008ta,Schumann:2007mg} using the NNPDF3.0NNLO PDF set~\cite{ATL-PHYS-PUB-2016-003}.  The matrix element calculation was performed with \textsc{Comix}~\cite{Gleisberg:2008fv} and \textsc{OpenLoops}~\cite{Cascioli:2011va} and was matched using the MEPS@NLO prescription~\cite{Hoeche:2012yf}.
 
Several subleading backgrounds were also simulated. The diboson~+~jets samples were generated using \SHERPA~2.1.1~\cite{multibosonMC} and the CT10 PDF set. Associated production of $\ttbar \Wboson$ and $\ttbar \Zboson$ ($\ttbar V$) were generated with an NLO matrix element using {\amcatnlo} interfaced to \pythia~8.210 and the NNPDF3.0NNLO PDF set. Samples of $\Wboson t$ single-top-quark backgrounds were generated with \POWHEGBOXV{v1} at NLO accuracy using the CT10 PDF set. The production of four top quarks ($t\bar{t}t\bar{t}$) and $t\bar{t}WW$ was simulated with \amcatnlo{} at LO accuracy and interfaced to \pythia~8.186. Background sources with non-prompt leptons contribute negligibly to this search.
 
Multijet samples are used to compare the data identification efficiency of the $\abb$ decays with simulation. These samples were generated using \pythia~8.186, with the LO NNPDF2.3 PDF set and the A14 tune. To increase the number of simulated events with semileptonically decaying hadrons used in this analysis, samples of multijet events filtered to have at least one muon with \pT\ above 3~\GeV\ and $|\eta| < 2.8$ were produced with $\pythia$ using the same version, PDF set, and underlying-event tunes as the unfiltered multijet samples. Both the filtered and unfiltered multijet samples produced with $\pythia$ were processed through the same ATLAS detector simulation.
 
\section{Object reconstruction and selection}
\label{sec:objreco}
 
This search relies on the efficient reconstruction of electrons and muons in order to identify leptonically decaying $\Zboson$ bosons and the reconstruction of jets to identify $\abb$ decays.
 
Electrons are reconstructed from energy deposited in clusters of cells in the electromagnetic calorimeter matched to tracks in the
inner detector~\cite{PERF-2017-01} and are required to have $\pt>15~{\GeV}$ and $|\eta|<2.47$. Candidates in the transition region between the barrel and  endcap calorimeters, $1.37<|\eta|<1.52$, are excluded. Electrons are identified using the ``Tight'' criterion based on a likelihood discriminant~\cite{ATLAS-CONF-2016-024}. Muons are reconstructed by combining matching tracks in the inner detector and the muon spectrometer, and are required to have $\pt>10~{\GeV}$ and $|\eta|<2.4$. Muon candidates must satisfy the ``Medium'' identification criterion~\cite{PERF-2015-10}. An isolation requirement based on the momentum of the tracks and the calorimeter energy around each lepton candidate is imposed to distinguish between leptons coming from the decay of a $\Zboson$ boson and those from non-prompt sources~\cite{ATLAS-CONF-2016-024,PERF-2015-10}. Additionally, all lepton candidates are required to be consistent with the primary vertex, chosen as the reconstructed vertex with the highest sum of the $\pt^2$ of its associated tracks.
 
Jets are reconstructed from three-dimensional topological energy clusters~\cite{PERF-2014-07} in the calorimeter
using the \antikt\ jet algorithm~\cite{Cacciari:2008gp} implemented in the FastJet package~\cite{Cacciari:2011ma} with a radius parameter of 0.4. Jets are calibrated using energy- and $\eta$-dependent corrections~\cite{PERF-2016-04} and are required to have $\pt >20~{\GeV}$ and $|\eta|<2.5$. Events containing jets arising from non-collision sources or detector noise are removed~\cite{ATLAS-CONF-2015-029}.  Finally, a track-based criterion, the jet vertex tagger (JVT), is used to reduce contributions from jets arising from pileup~\cite{ATLAS-CONF-2014-018}. In the region $|\eta|<2.5$, jets are tagged as containing $b$-hadrons using a multivariate discriminant (MV2) score~\cite{b-tagging2016}. The MV2 score is obtained from a boosted decision tree (BDT) that combines several algorithms that identify tracks with large impact parameters, secondary vertices, and the topological structure of weak $b$- and $c$-hadron decays inside jets. The BDT was trained using jets reconstructed with the \antikt\ algorithm with a radius parameter $R=0.4$ from $\ttbar$ simulated events to discriminate $b$-jets from $c$-jets and light-flavor jets~\cite{PERF-2012-04}. In this search, the same BDT is used with a novel strategy described in Section~\ref{subsec:recoabb}.
 
\section{Identification of low-mass resonances decaying into $b$-quark pairs}
\label{sec:atobb_id}
 
\subsection{Reconstruction and identification of $\abb$ decays}
\label{subsec:recoabb}
 
For low-mass $a$-bosons, the $b$-quarks from $a$-boson decay tend to have small angular separation $\Delta R$ and can be reconstructed either as a single jet or as multiple jets in the calorimeter depending on their angular separation and the clustering algorithm used. In order to include both cases, all calibrated jets reconstructed using the \antikt\ jet algorithm with radius parameter $R=0.4$ and $\pt > 20\,\GeV$ are clustered again, using an \antikt\ algorithm with radius parameter $R=0.8$~\cite{Nachman:2014kla}. The radius parameter was chosen to optimize the signal acceptance in the mass range considered. Each $R=0.8$ jet is considered as a reconstructed $\abb$ candidate. The $R=0.8$ jet will often contain a single \antikt\ constituent jet with radius parameter $R=0.4$ when the angular separation $\Delta R$ between the $b$-quarks from the $\abb$ decay is less than $0.4$. The four-momentum of an $\abb$ candidate is the sum of all four-momenta of the set of constituent $R=0.4$ jets. Since the $R=0.4$ constituent jets are calibrated, no additional momentum calibration is necessary. 
 
The hadronization of the two $b$-quarks which come from an $a$-boson decay is identified using variables sensitive to the number of $b$-hadrons and the mass of the $a$-boson. The values of these variables are calculated using tracks with $\pt>0.4~\GeV$ matched to the reconstructed $\abb$ candidate. The matching is performed using the ghost-association method~\cite{Cacciari:2007fd}, which treats the tracks as four-vectors of infinitesimal magnitude during the jet reconstruction and assigns them to the $\abb$ candidate with which they are clustered. Tracks from the hadronization of different $b$-quarks are separated by splitting the set of tracks matched to an $\abb$ candidate into multiple track-jets. Ideally, the decay of each $b$-quark should be associated with a different track-jet. In this search, the track-jets are reconstructed by clustering all matched tracks using an exclusive-$k_{t}$ algorithm that produces either two (Ex$k_{t}^{(2)}$) or three (Ex$k_{t}^{(3)}$) final jets~\cite{Catani:1993hr}. The exclusive-$k_{t}$ algorithm implements a sequential clustering in which the two tracks with the smallest $k_{t}$-distance, defined as the product of the minimum $\pt$ of the two tracks and their distance $\Delta R$, are clustered together if this distance is smaller than the transverse momentum of all tracks. If two tracks are clustered together, their momenta is summed and the two considered as a single object in the next iteration of the sequential clustering. If the transverse momentum of a track is smaller than all $k_t$-distances, the track is discarded. Tracks clustered together are considered a final-state track-jet. The sequential clustering is interrupted after the step in which all the tracks have been clustered in the desired number of final-state track-jets~\cite{Salam:2009jx}. The splitting into three final jets attempts to capture cases where significant additional radiation is present. The strategy presented here to identify the two $b$-quark flight directions as different track-jets differs from the method documented in Ref.~\cite{Aad:2019uoz}, where the inclusive version of the \antikt\ algorithm is used. At low $a$-boson momenta, the exclusive-$k_t$ algorithm is able to identify the two $b$-quark flight directions in separate track-jets more often than the inclusive \antikt\ algorithm. For a simulated signal event sample with $m_a=20\,\GeV$, the inclusive \antikt\ algorithm associates the $b$-quark flight directions with different track-jets in 46\% of cases. In contrast, the flight directions are associated with different exclusive-$k_t$ track-jets in nearly 100\% of cases.
 
The variables used for the $\abb$ identification are calculated using the exclusive-$k_t$ track-jets. For the track-jets calculated with the Ex$k_{t}^{(2)}$ algorithm, the variables used are the MV2 scores of the two track-jets, as well as their angular separation $\Delta R$ and their $\pt$ asymmetry, defined as $(\pt^1-\pt^2)/(\pt^1+\pt^2)$. For Ex$k_{t}^{(3)}$ track-jets, the same variables are used, but they are calculated with the two track-jets with highest and lowest MV2 scores among the three track-jets. The eight variables are used simultaneously. The MV2 scores identify the presence of a $b$-hadron in the track-jets. Track-jets with large $\Delta R$ separation occur in the decay of a massive state. Track-jets with very large $\pt$ asymmetry can arise from final-state radiation. The variables calculated with Ex$k_{t}^{(2)}$ track-jets provide most of the discriminating power between signal and background, while the variables calculated in Ex$k_{t}^{(3)}$ help disentangle cases where Ex$k_{t}^{(2)}$ would fail to identify the flight direction of the $\abb$ decay products.
 
\begin{figure*}[htbp]
\begin{center}
\subfloat[]{\includegraphics[width=0.49\textwidth]{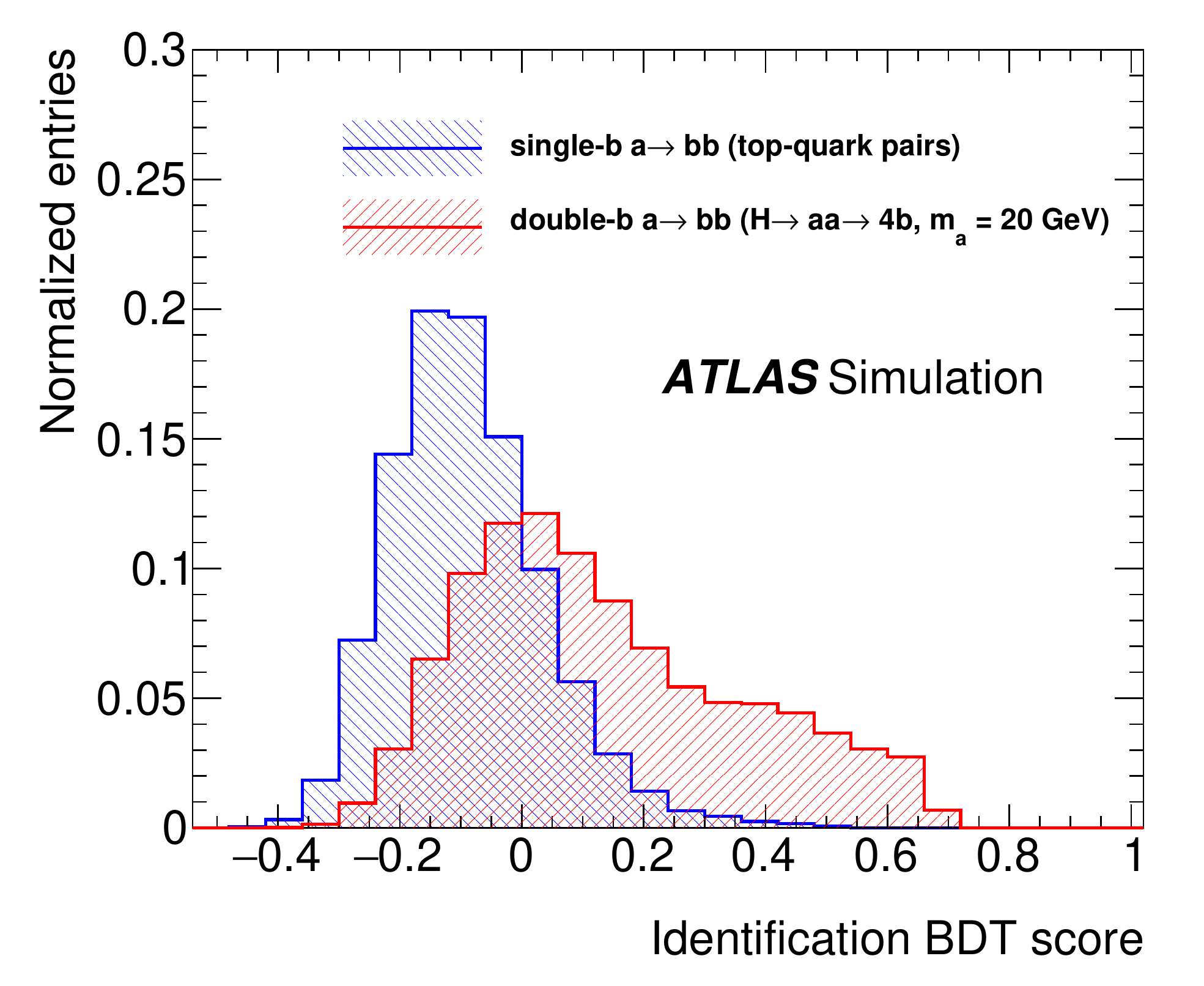}}
\subfloat[]{\includegraphics[width=0.51\textwidth]{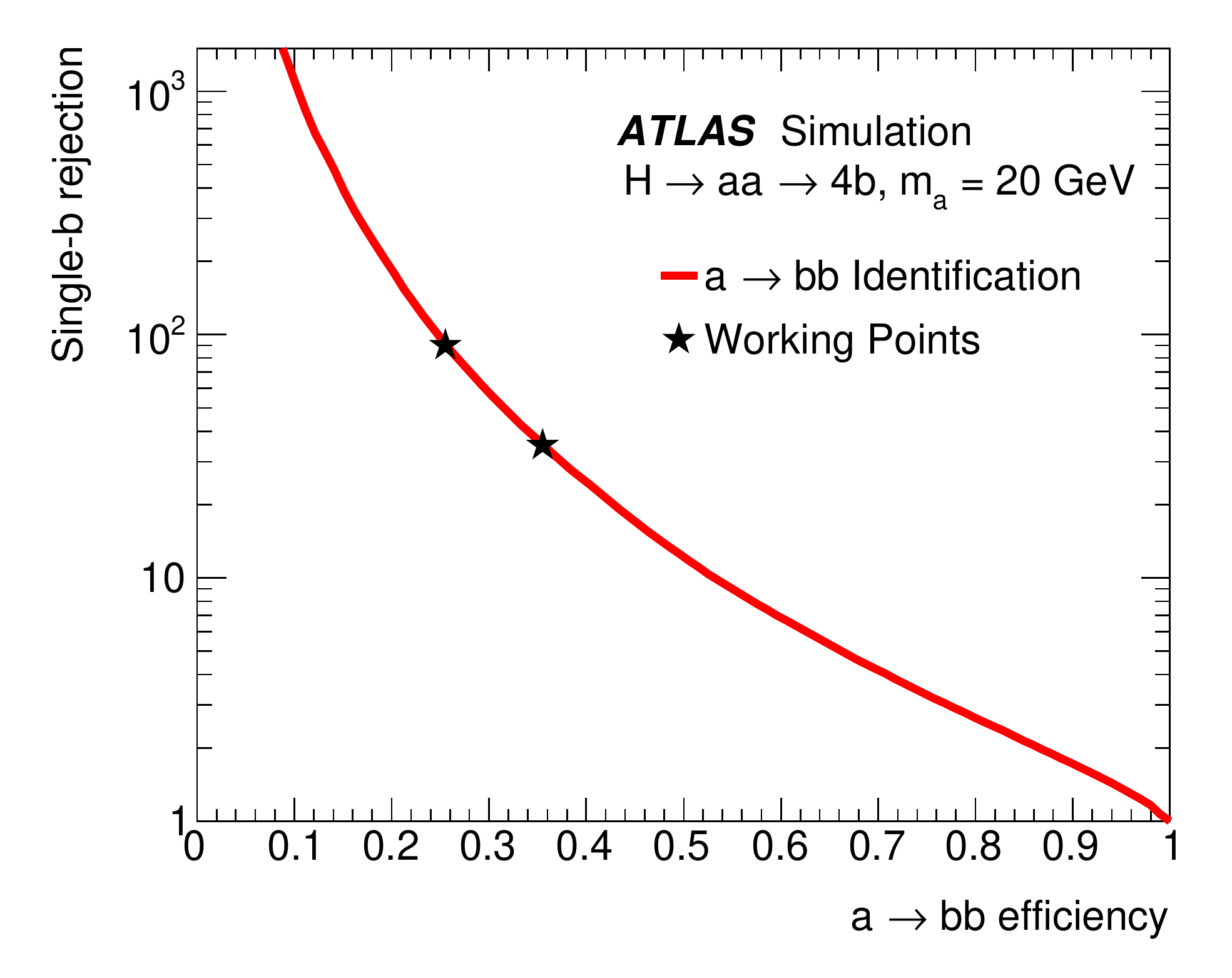}}\\
\caption{(a) Identification BDT score distributions for signal and background $\abb$ candidates and (b) signal efficiency as a function of the inverse of the $\ttbar$ background efficiency (rejection). For the signal $H\to aa\to (b\bar{b})(b\bar{b})$ sample with $m_{a}=20\,\text{GeV}$, both $b$-quarks are required to lie within the reconstructed candidate, while for the background $\ttbar$ sample the reconstructed candidate contains a single $b$-quark.
In Figure~(b), the left and right stars indicate the tight and the loose WPs, respectively, which are used to define, as described in the text, the HPC and LPC.}
\label{fig:roc}
\end{center}
\end{figure*}

A BDT is trained with these variables to obtain an efficient identification criterion that distinguishes $\abb$ candidates in signal events that have two $b$-quarks produced in the decay of a low-mass resonance, from those in top-quark pair events that contain a single $b$-quark decay. A sample of simulated SM $\ttbar$ events is used as a source of $\abb$ candidates with a single $b$-quark decay, while a simulated signal event sample with $m_a=20\,\GeV$ is used as a source of $\abb$ candidates with two $b$-quark decays. The transverse momentum and angular distributions are not included as inputs for the BDT training, but the differences in these distributions among signal and background are partially taken into account since they are correlated with other variables. In order to classify the $b$-quark multiplicity of an $\abb$ candidate, $b$-hadrons in the simulation of the $b$-quark hadronization with $\pt > 5\,\GeV$ are matched to the candidates using the same ghost-association method as described above. Figure~\ref{fig:roc} shows the predicted score and efficiency for signal and background events using the trained BDT. The BDT discriminator is also efficient in rejecting events without $b$-quarks, even if such sample was not explicitly included in the BDT training. Two event categories based on the BDT score are defined for the analysis using a tight and a loose working point (WP). A high-purity category (HPC) for $\abb$ candidates is selected by requiring a BDT score larger than the tight WP, while a low-purity category (LPC) is selected from candidates with BDT score between the loose and the tight WPs. The tight WP is defined by a BDT score of 0.3 while the loose WP is defined by a BDT score of 0.1. The tight WP is chosen such that it provides a background rejection $1/100$ in order to reduce the backgrounds from $\Zjets$ and $\ttbar$ events. The LPC contains a relatively large number of events from processes with zero or one $b$-quark and is used to select background-enriched samples in the search. Reconstructed $\abb$ candidates in the LPC and HPC are defined as identified $\abb$ candidates and are used in this search. The signal efficiency of the two WPs vary with the mass of the $a$-boson since mass-dependent variables are used in the training.
 
The efficiency of the $\abb$ identification is measured in data by selecting a multijet sample enriched in gluon decays into $b$-quarks, $g\to \bbbar$. In order to measure the efficiency of the identification criterion for both signal and background, $\abb$ candidates are categorized according to the flavor of the track-jets that are reconstructed with the Ex$k_{t}^{(2)}$ algorithm, while the Ex$k_{t}^{(3)}$ track-jets are used exclusively for identification purposes. All $b$- and $c$-hadrons present in the event simulation with $\pt > 5$~GeV are matched to the track-jets using the ghost-association method. The track-jets are assigned different flavor tags B, C, or L (light flavor) as follows. If a track-jet has at least one simulated $b$-hadron matched to it, it is classified as B. If it does not contain a simulated $b$-hadron, but has a simulated $c$-hadron matched to it, it is classified as C. Otherwise it is classified as L. The flavor of an $\abb$ candidate is determined by the flavor of the two Ex$k_{t}^{(2)}$ jets. Most signal $\abb$ candidates are $\text{BB}$ candidates, while most background candidates are BL candidates. A signal candidate can be classified as BL when the two $b$-quarks decay inside the same track-jet or when they have $\pt \leq 5\,\GeV$. The identification efficiencies for $\text{BB}$ and $\text{BL}$ $\abb$ candidates are measured separately in data for three transverse momentum ranges: $30\,\GeV\,\leq p^{\abb}_{\text{T}} < 90\,\GeV$, $90\,\GeV\,\leq p^{\abb}_{\text{T}} < 140\,\GeV$, and $p^{\abb}_{\text{T}}  \geq 140\,\GeV$. These three ranges were chosen based on the $p^{\abb}_{\text{T}}$ spectrum in signal samples and on the number of events in the multijet data sample used for the efficiency measurement. The complete procedure described below is applied independently in each transverse momentum range.
 
\subsection{Efficiency measurement of $\abb$ identification}
\label{subsec:effmeas}
 
The strategy for the efficiency measurement in data closely follows that used in the measurement of the identification efficiency for boosted $125~\GeV$ Higgs boson decays into a pair of $b$-quarks~\cite{Aad:2019uoz}. A multijet sample is selected from a suite of single-jet triggers that differ by their jet $p_{\text{T}}$ threshold. Only a small fraction of the events identified by the triggers with low $p_{\text{T}}$ threshold are recorded.
The choice of which jet events to keep is random and results in an effective integrated luminosity smaller than the total recorded by the ATLAS experiment, but does not introduce any selection bias. The fraction of events kept is known as the trigger prescale fraction. Triggers with prescale fraction less than one are called prescaled triggers. The lowest jet $p_{\text{T}}$ threshold for which all events are kept is $300\,\GeV$. When comparing events recorded with prescaled and unprescaled triggers, each event is weighted by the inverse of the prescale fraction of the corresponding trigger used to record it.
 
The $\abb$ reconstruction described in Section~\ref{subsec:recoabb} is applied to the multijet sample. The events recorded by the multijet triggers are
dominated by $\text{LL}$ candidates. Since muons are often produced in semileptonic decays of $b$-hadrons, a sample with a larger fraction of $\text{BB}$ and $\text{BL}$ candidates is selected by requiring exactly one muon matched to one of the Ex$k_{t}^{(2)}$ track-jets. The track-jet matched to the muon is called the muon-matched track-jet, while the other one is called the non-muon-matched track-jet. The selected events are compared with simulated multijet events. In order to account for possible mismodeling of the flavor fractions in simulation relative to those in data, a correction is applied to the simulated event sample. The correction is described in detail in Ref.~\cite{Aad:2019uoz} and only a brief summary is given here. The simulated jet sample is split into subsamples depending on the flavor classification of the $\abb$ candidate: $\text{BB}$, $\text{BL}$, $\text{CC}$, $\text{CL}$, and $\text{LL}$. The selected $\text{BC}$ fraction in the multijet sample is negligible. The fraction of each subsample is corrected by fitting the distribution of the signed transverse impact parameter significance $S_{d0}^{\text{jet}} = d_0 /\sigma(d_0)$ of the two Ex$k_{t}^{(2)}$ track-jets to data. The $S_{d0}^{\text{jet}}$ of a track-jet is defined as the average of the three largest signed transverse impact parameter significances $S_{d0}^{\text{trk}}$ of its constituent tracks, since this observable is used to identify the long lifetime of $b$- and $c$-hadrons. The average is used to minimize the impact of misreconstructed tracks on this observable. The track impact parameter $d_0^{\text{trk}}$ is calculated using the vector from the primary vertex to the point of closest approach of the track. The absolute value of $d_0^{\text{trk}}$ is the norm of the projection of this vector in the transverse plane, while the sign depends on the angle between this vector and the track-jet $\vec{\pt}$. If this angle is less than $\pi/2$,  $d_0^{\text{trk}}$ is taken as positive. For angles larger than $\pi/2$, the track impact parameter is considered negative. Large negative impact parameters are often obtained from interactions with the detector material and not from a long-lived $b$- or $c$-hadron decay, since the direction of the decay is not correlated with the jet axis.
 
\begin{figure*}[htbp]
\begin{center}
\subfloat[]{\includegraphics[width=0.5\textwidth]{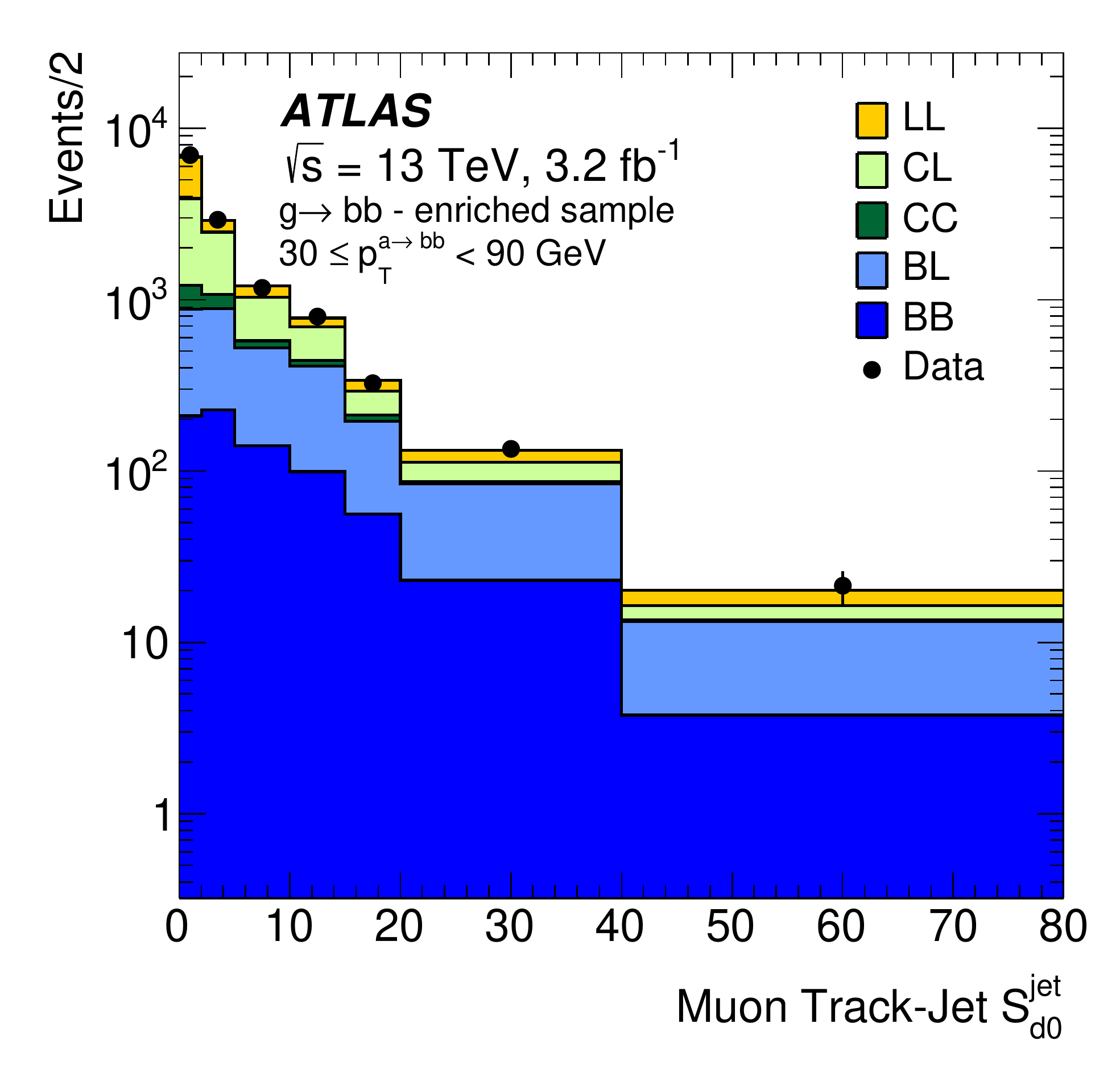}}
\subfloat[]{\includegraphics[width=0.5\textwidth]{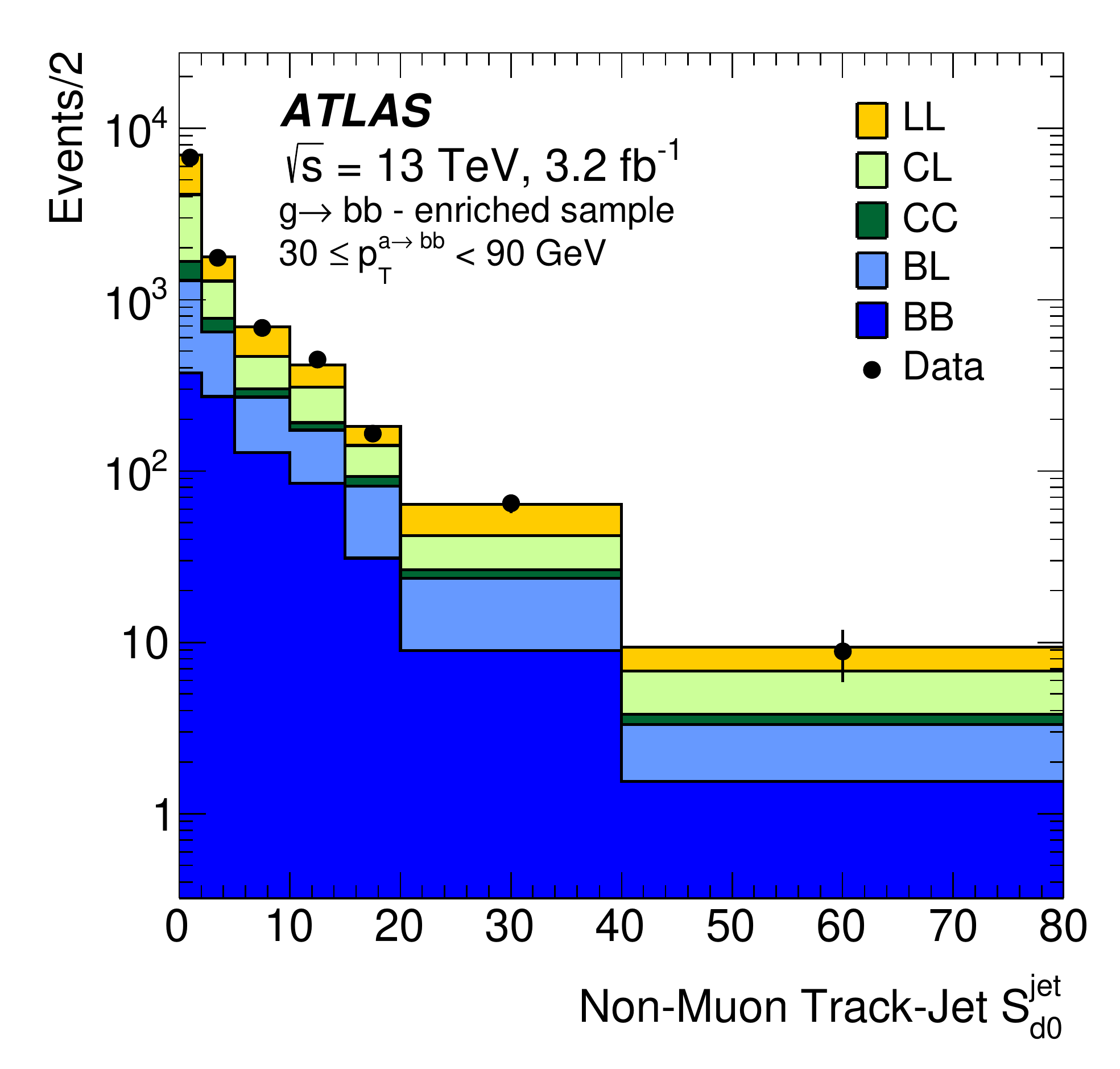}}\\
\caption{Averaged signed impact parameter significance $S_{d0}^{\text{jet}}$ distributions of the track-jet (a) with a muon inside and (b) without a muon, after performing the fit of the flavor fractions to data. The total simulation yield is scaled to the same number of events observed in data. The fit is performed separately in $p^{\abb}_{\text{T}}$ ranges. The figure shows the $30\,\GeV\,\leq p^{\abb}_{\text{T}} < 90\,\GeV$ range of $\abb$ candidates.}
\label{fig:flavorfit}
\end{center}
\end{figure*}

A total of four flavor correction factors that scale the $\text{BB}$, $\text{BL}$, $\text{CC}$, and $\text{CL}$ subsamples are determined from a Poisson likelihood fit to data. The scale parameter for the $\text{LL}$ subsample is determined implicitly by requiring that the total number of candidates in simulation is the same as in data. The covariance matrix of these four parameters is determined from the statistical uncertainties and correlations, but also from the limited knowledge of the jet energy scale, and from the uncertainty in the impact parameter resolution following the method described in Ref.~\cite{Aad:2019uoz}. Figure~\ref{fig:flavorfit} shows the result of this fit to data for the transverse momentum range $30\,\GeV\,\leq p^{\abb}_{\text{T}} < 90\,\GeV$.
 
\begin{figure*}[htbp]
\begin{center}
\subfloat[]{\includegraphics[width=0.5\textwidth]{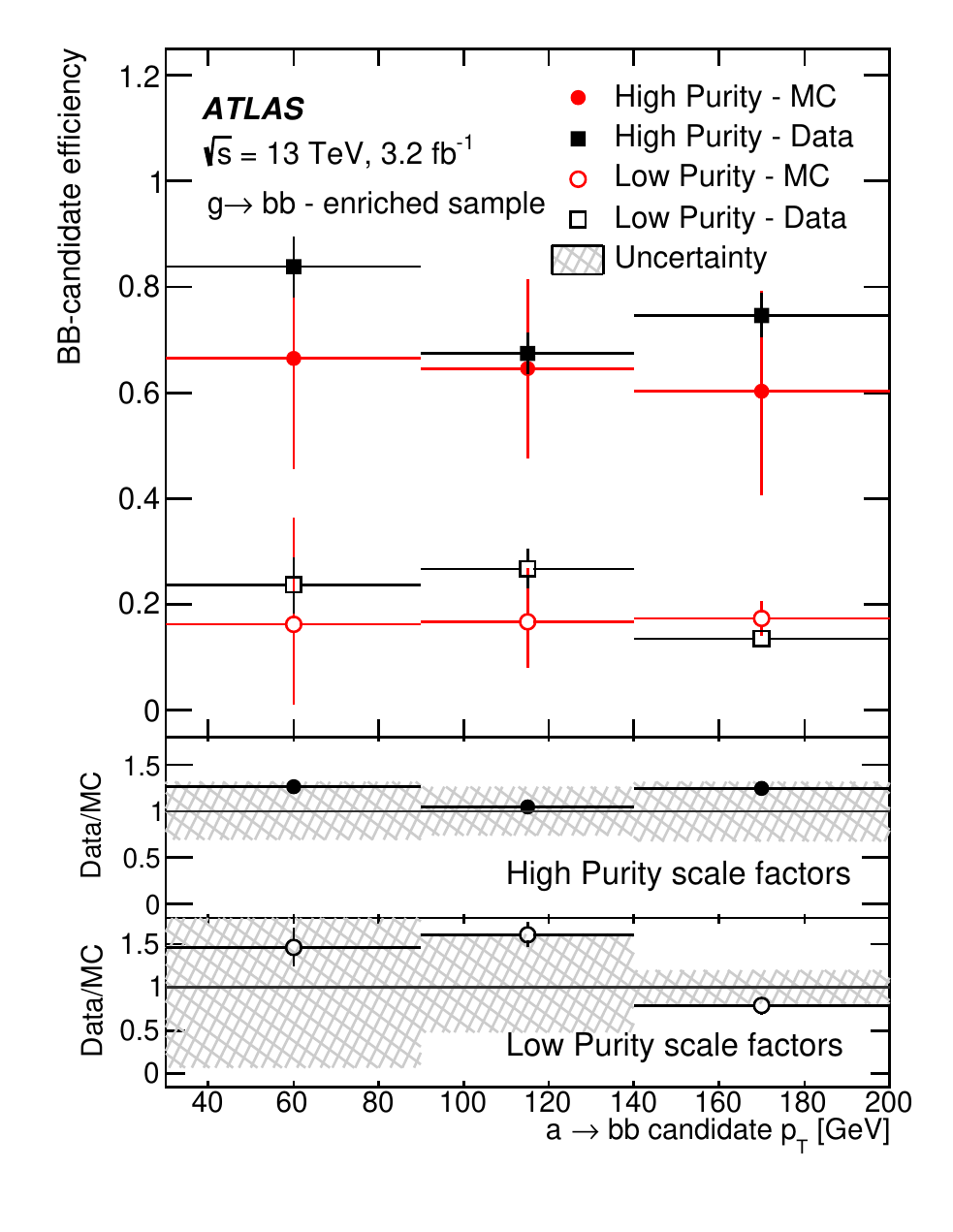}}
\subfloat[]{\includegraphics[width=0.5\textwidth]{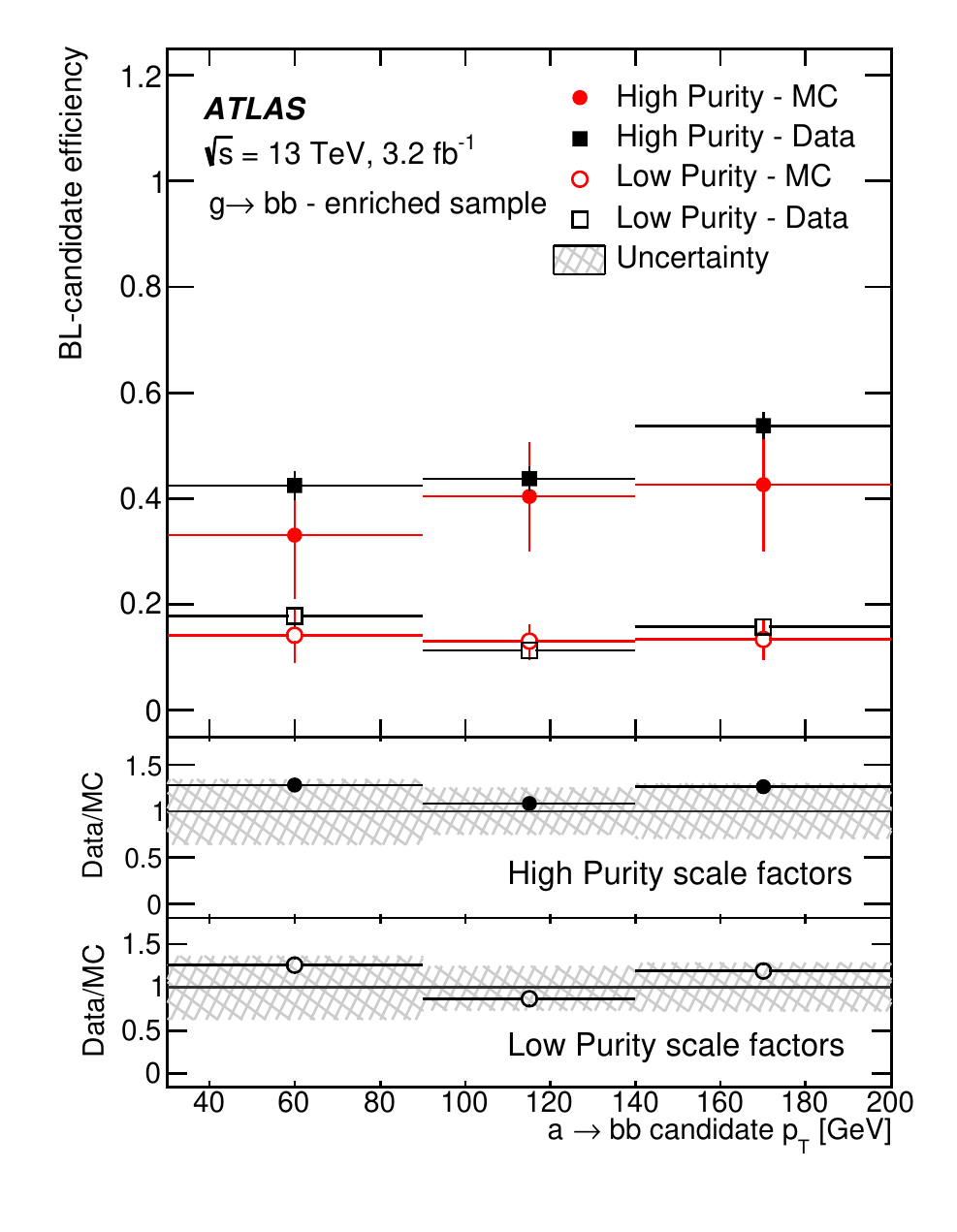}}\\
\caption{Efficiency of the $\abb$ identification criteria measured in data and simulated multijet events. The efficiency is measured in three transverse momentum ranges, separately for (a) $\text{BB}$ and (b) $\text{BL}$ candidates in both the HPC and LPC. The ratio of the measured values in data and simulation (bottom panels) are SFs used in the analysis when comparing simulation with data. The error bars in the top panels are statistical only, while the hashed bands on the ratios in the bottom panels include the full systematic uncertainties.}
\label{fig:datamceff}
\end{center}
\end{figure*}

After the flavor correction is applied, the $\abb$ identification BDT is used to select events in both the HPC and LPC. Once the identification criteria are used, only the $\text{BB}$ and the $\text{BL}$ subsamples contribute significantly. Any residual disagreement in these regions is the result of a difference in the $\abb$ identification efficiency between data and simulation. A scale factor (SF) is defined as the ratio of the two efficiencies, $\text{SF} = \varepsilon_{\text{DATA}}/\varepsilon_{\text{MC}}$, for each flavor subsample. Only the $\text{BB}$ and $\text{BL}$ SFs are measured for both HPC and LPC. All other flavors are subleading after applying the identification criterion, and for these the efficiency in data is considered the same as in simulation. In order to measure the $\text{BB}$ and $\text{BL}$ SFs, in both the HPC and LPC, a second Poisson likelihood fit of the $S_{d0}^{\text{jet}}$ distribution to data is performed after using the identification BDT to select events in both simulation and data. The four SFs measured in each of the three $\pt^{\abb}$ ranges constitute 12 parameters in total. The complete list of uncertainties is described in Section~\ref{subsec:systabb}. Figure~\ref{fig:datamceff} shows the measured efficiencies in both data and simulation, for $\text{BB}$ and $\text{BL}$ candidates. The bottom panel in the same figure shows the SF as defined above.
 
\subsection{Systematic uncertainties in the $\abb$ identification}
\label{subsec:systabb}

Several sources of uncertainty are considered when building the covariance matrix of the 12 SFs. The statistical
uncertainties and correlations are interpreted directly from the likelihood fit to data. The impact of systematic uncertainties is considered by varying the appropriate quantity in the simulated event samples within $\pm 1\sigma$ for each source separately. The impact of each systematic uncertainty is assessed as the difference in the measured SF when fitting the nominal sample and the one with the corresponding source variation. The covariance matrix from the four flavor-fraction corrections described in Section~\ref{subsec:effmeas} is propagated to the SF covariance matrix. The impact of the limited knowledge of the jet energy scale is, once again, considered in the covariance matrix. The uncertainty arising from the choice of hadronization model is included through its effect on the MV2 scores and propagated to the SF covariance matrix. This uncertainty changes the MV2 score by 5--10\% depending on the track-jet $\pt$~\cite{ATL-PHYS-PUB-2020-009} and has a minor impact on the SFs.
 
\begin{figure*}[htbp]
\begin{center}
\subfloat[]{\includegraphics[width=0.5\textwidth]{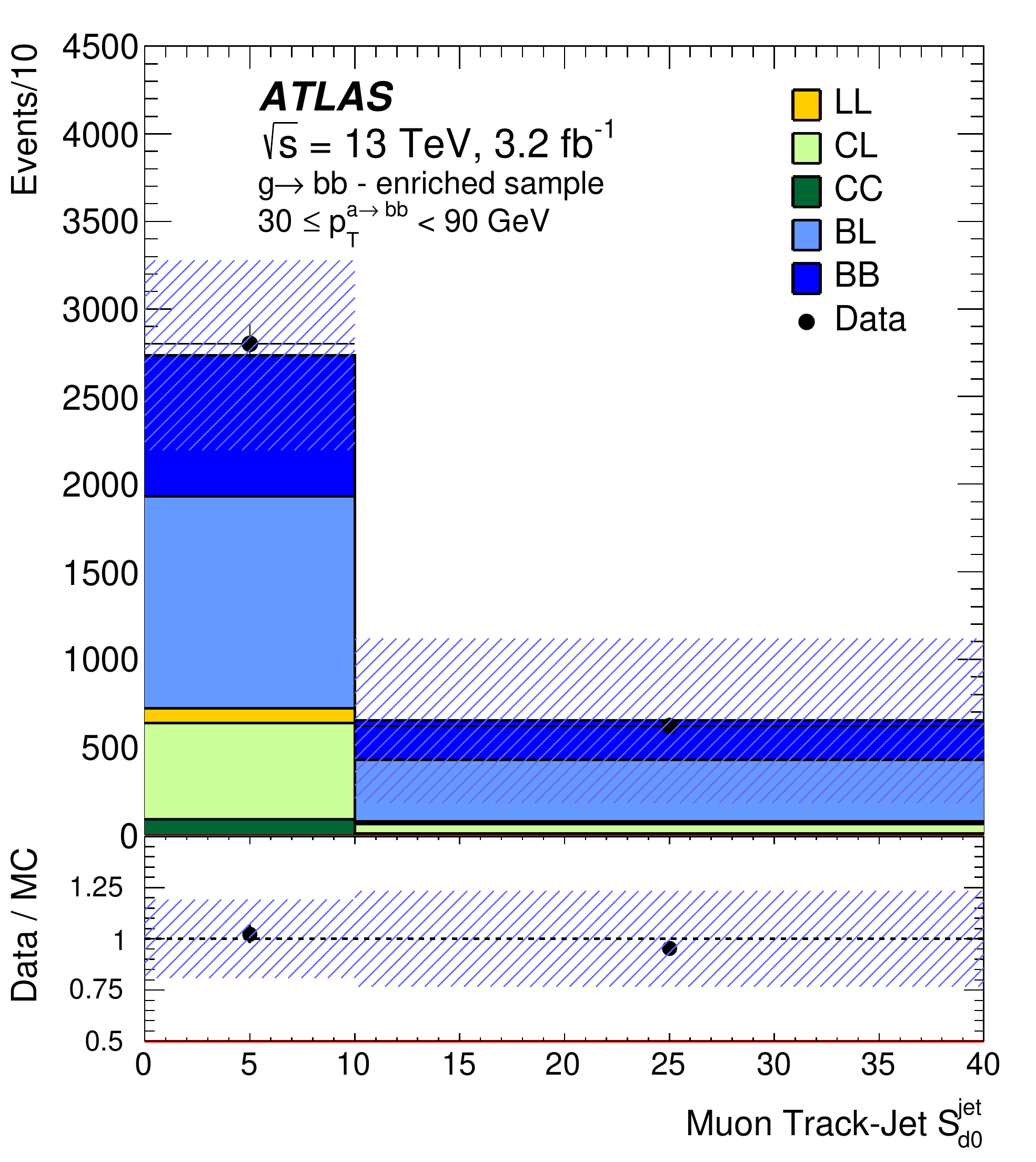}}
\subfloat[]{\includegraphics[width=0.5\textwidth]{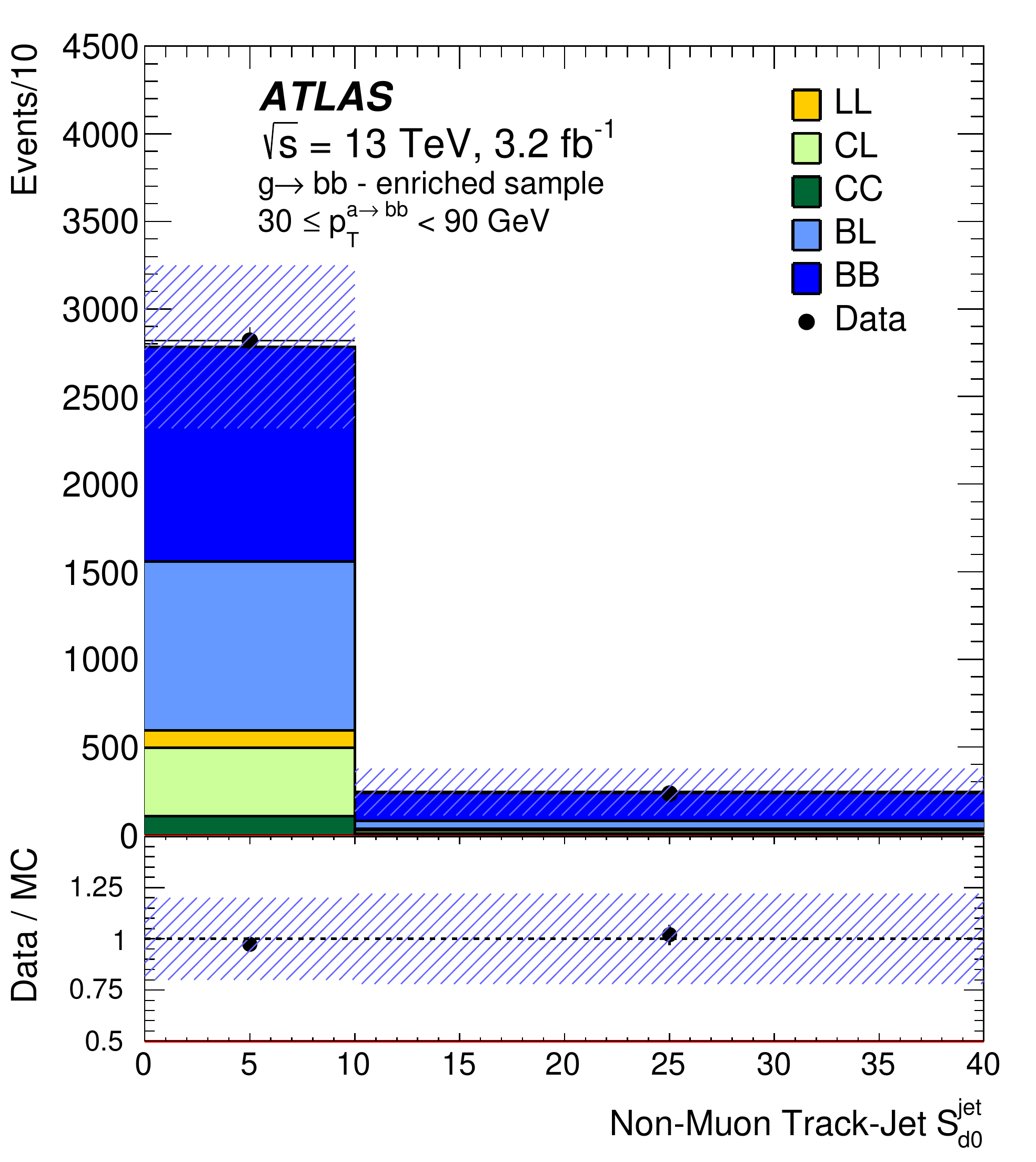}}\\
\caption{Averaged signed impact parameter significance $S_{d0}^{\text{jet}}$ distributions of the track-jet (a) with a muon inside and (b) without a muon, in the $30\,\GeV\,\leq p^{\abb}_{\text{T}} < 90\,\GeV$ range of $\abb$ candidates in the HPC. The hashed area represents the total uncertainty in the predicted yields.}
\label{fig:sffit}
\end{center}
\end{figure*}

Two additional sources of uncertainty are considered. First, there is a possible mismodeling of the efficiency for candidates with flavor other than $\text{BB}$ or $\text{BL}$. These components are highly suppressed by the BDT selection. An uncertainty of $50\%$ in the efficiency of other flavor components is propagated to the covariance matrix, with negligible impact. The chosen value of $50\%$ is based on the level of agreement between the $S_{d0}^{\text{jet}}$ distribution in data and simulation, after the BDT criteria are applied, but before the SF fit. Second, there is a possible bias from the selection of $\abb$ candidates with muons. In order to assess it, the measurement is repeated by selecting $\abb$ candidates with two muons, one inside each track-jet, and comparing the result with the one obtained above. The sample with two muons contains a negligible number of $\text{BL}$ candidates and it is only possible to measure the $\text{BB}$ SFs. The difference between the SF measured with the one-muon sample and the one measured with the two-muon sample is taken as an estimate of the systematic uncertainty due to a possible bias in the procedure. The same uncertainty is applied to the $\text{BL}$ SFs. This uncertainty varies in each $p_{\text{T}}^{\abb}$ range, but it is approximately $20\%$ and is the dominant uncertainty in the $\pt$ ranges with a large number of candidates. Figure~\ref{fig:sffit} shows the $S_{d0}^{\text{jet}}$ distribution in the $30\,\GeV\,\leq p^{\abb}_{\text{T}} < 90\,\GeV$ range, for the HPC, after fitting for both the $\text{BB}$ and $\text{BL}$ SFs and including all the uncertainties described here.
 
\begin{figure*}[htbp]
\begin{center}
\includegraphics[width=0.5\textwidth]{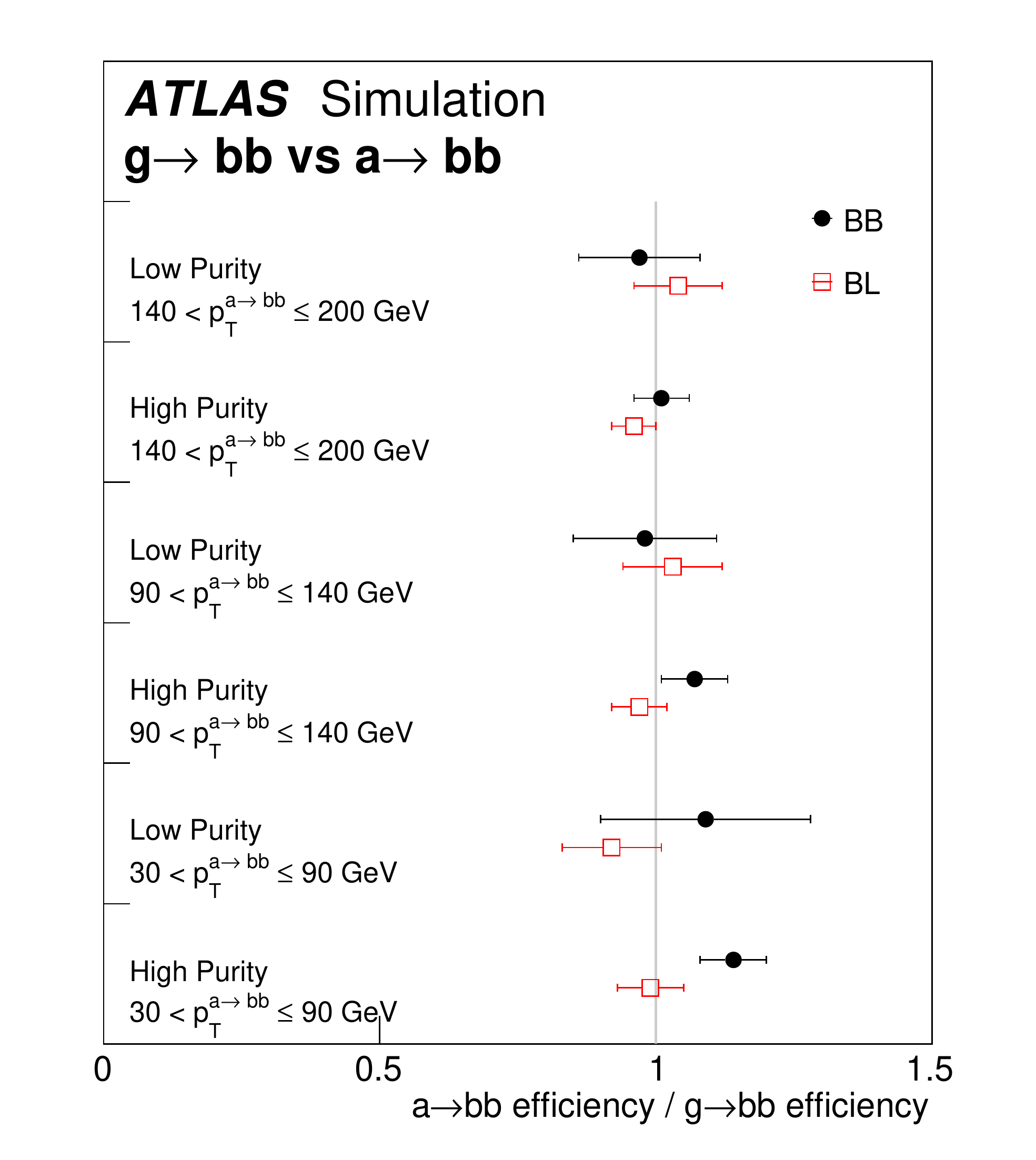}
\caption{Efficiency measured using a simulated multijet pseudo-data where $g\to \bbbar$ decays are replaced by $\abb$ decays with mass $m_a = 20\,\text{GeV}$. The efficiency is measured separately for $\text{BB}$ and $\text{BL}$ samples in the same $\pt$ ranges used in the data-to-simulation SF measurement. The values can be interpreted as the ratio between the SFs for a particle with mass $m_a=20\,\text{GeV}$ and the ones for a massless gluon. Only the statistical uncertainties are indicated.
}
\label{fig:efftransfer}
\end{center}
\end{figure*}

These SFs are used when comparing simulated signal and background events with data. For the selected background events, the distributions of the variables used for the $\abb$ identification are similar to the ones in $g\to \bbbar$ events. However, for the signal, due to the nonzero mass of the $a$-boson, the distributions can be quite different, especially for the variables that are sensitive to the mass of the particle. The method presented here relies on the fact that any residual disagreement accounted for by the SFs is independent of the $a$-boson mass. In order to test this hypothesis, the efficiency measurement is repeated replacing data with a pseudo-data built using the same multijet simulated sample used to obtain the $S_{d0}^{\text{jet}}$ templates but where gluons were replaced by a spin-0 $a$ boson with mass $m_a = 20\,\text{GeV}$ before the decay to two $b$ quarks. Figure~\ref{fig:efftransfer} shows the results of using this pseudo-data in each of the categories considered above, which can be interpreted as the ratio between the SF for a particle with mass $m_a=20\,\text{GeV}$ and the one for a massless gluon. The overall distribution is consistent with unity within the statistical uncertainty of the simulated event sample.
 
\section{Analysis strategy}
\label{sec:analysis}
 
The analysis strategy targets events where a Higgs boson is produced in association with a $\Zboson$ boson. The candidate events are
required to be consistent with a $ZH$ event, where the $\Zboson$ boson decays into electrons or muons and the Higgs boson decay into two $a$-bosons each of which decay into a $b$-quark pair. Events are selected using triggers that require at least one electron
or muon. The event is further required to have two oppositely charged electrons or two oppositely charged muons and two reconstructed $\abb$ candidates. The leading electron or muon is required to have $p_{\text{T}} >
27\,\GeV$ and be matched to the lepton candidate reconstructed by the trigger
algorithms. The lepton momentum requirement and trigger matching are used so that all events have at least one lepton with $\pt$ above the trigger thresholds.
The dilepton mass must be consistent with the $\Zboson$ boson mass and is required to be in the range
$85\,\GeV\,< m_{\ell\ell} < 100\,\GeV$. Both $\abb$ candidates are required to satisfy
$p_{\text{T}} > 30\,\GeV$ and $|\eta| < 2.0$.
 
Two mass requirements are imposed to select events consistent with a
cascade decay $H\to aa\to (\bbbar)(\bbbar)$. Firstly, the mass difference $\Delta m^{\abb} = m^{a_1}-m^{a_2}$ between the
two $\abb$ candidates is required to satisfy $-25\,\GeV\,< \Delta
m^{\abb} < 25\,\GeV$. The ordering of $\abb$ candidates is based on their transverse momenta, with $a_1$
corresponding to the higher-$\pt$ $\abb$ candidate. Second, the mass of the pair of $\abb$
candidates is required to be consistent with the Higgs boson
mass. The compatibility is assessed with the reduced mass:
 
\begin{equation}\nonumber
m_{\text{red}} = (m^{a_1,a_2}-m_{H}) - (m^{a_1}+m^{a_2} - 2m_a),
\end{equation}
which probes the difference between the invariant mass of the two $\abb$ candidates, $m^{a_1,a_2}$, and the Higgs boson mass $m_H = 125\,\GeV$. It should be noted that $m_a$ is the mass hypothesis for the $a$-boson.
 
\begin{figure*}[htbp]
\begin{center}
\subfloat[]{\includegraphics[width=0.5\textwidth]{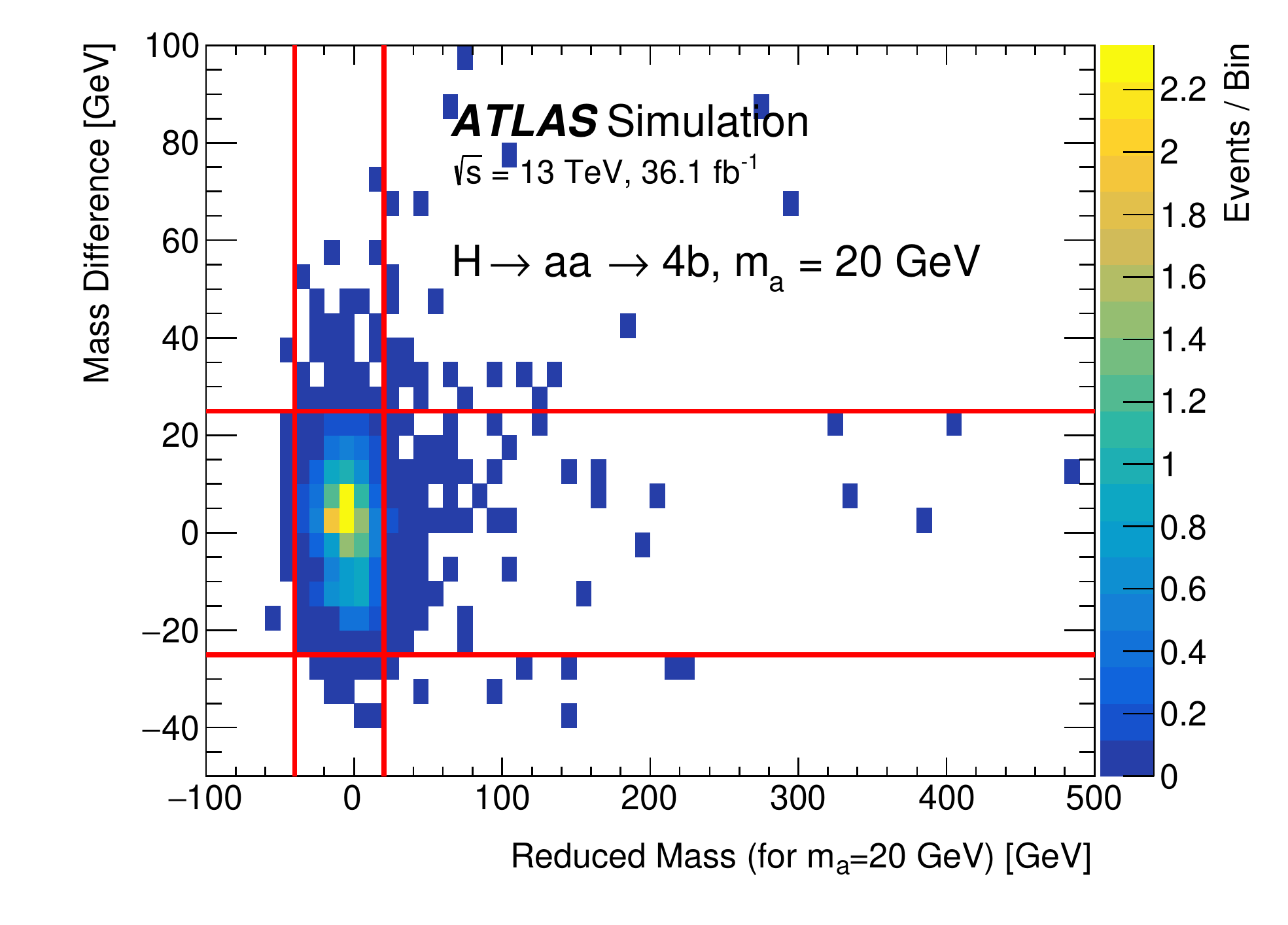}}
\subfloat[]{\includegraphics[width=0.5\textwidth]{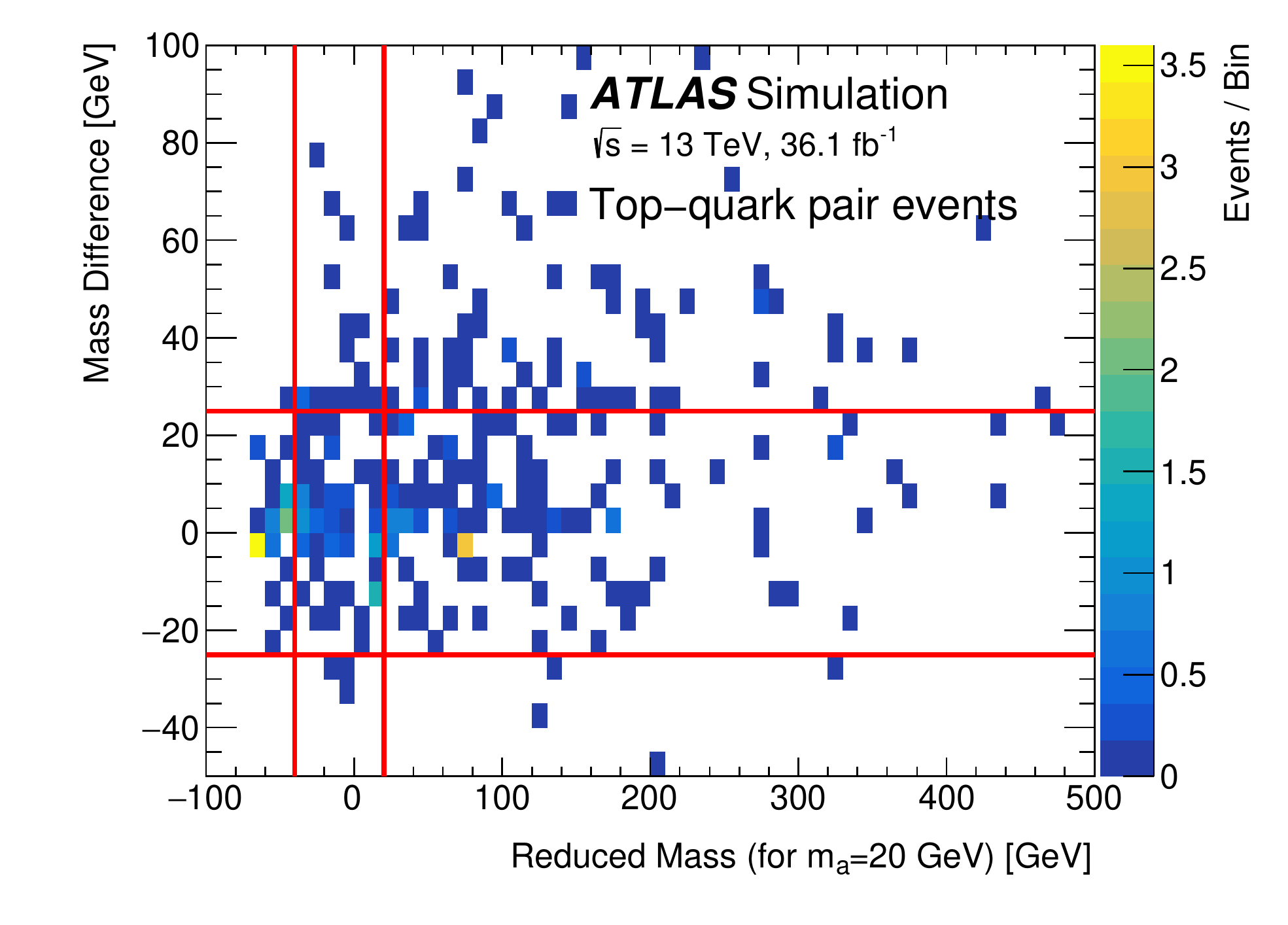}}\\
\caption{Distribution of (a) expected signal events and (b) top-quark pair background in a plane defined by the two mass requirements described in the text, $m_{\text{red}}$ and $\Delta m^{\abb}$. The mass requirements aim at selecting events where the two $\abb$ candidates have similar reconstructed masses and the mass of the pair of $\abb$ candidates is consistent with the Higgs boson mass. The signal events correspond to $ZH$, $H\to aa\to (\bbbar)(\bbbar)$ with $m_a = 20~\GeV$ and are normalized to the SM $pp \rightarrow ZH$ cross-section.
}
\label{fig:masswindow}
\end{center}
\end{figure*}

The reduced mass is required to satisfy $-40\,\GeV\,< m_{\text{red}} < 20\,\GeV$, ensuring that the selection is highly efficient. The presence of $m_a$ in the event selection means that different events are used to search for different mass hypotheses. No conditions on the individual values of $m^{a_1}$ and $m^{a_2}$ are imposed.
The selected mass window, as a function of mass difference and reduced mass, is shown in Figure~\ref{fig:masswindow} for signal events with $m_a=20\,\GeV$ and top-quark pair events.
 
\begin{figure*}[htbp]
\begin{center}
\includegraphics[width=0.55\textwidth]{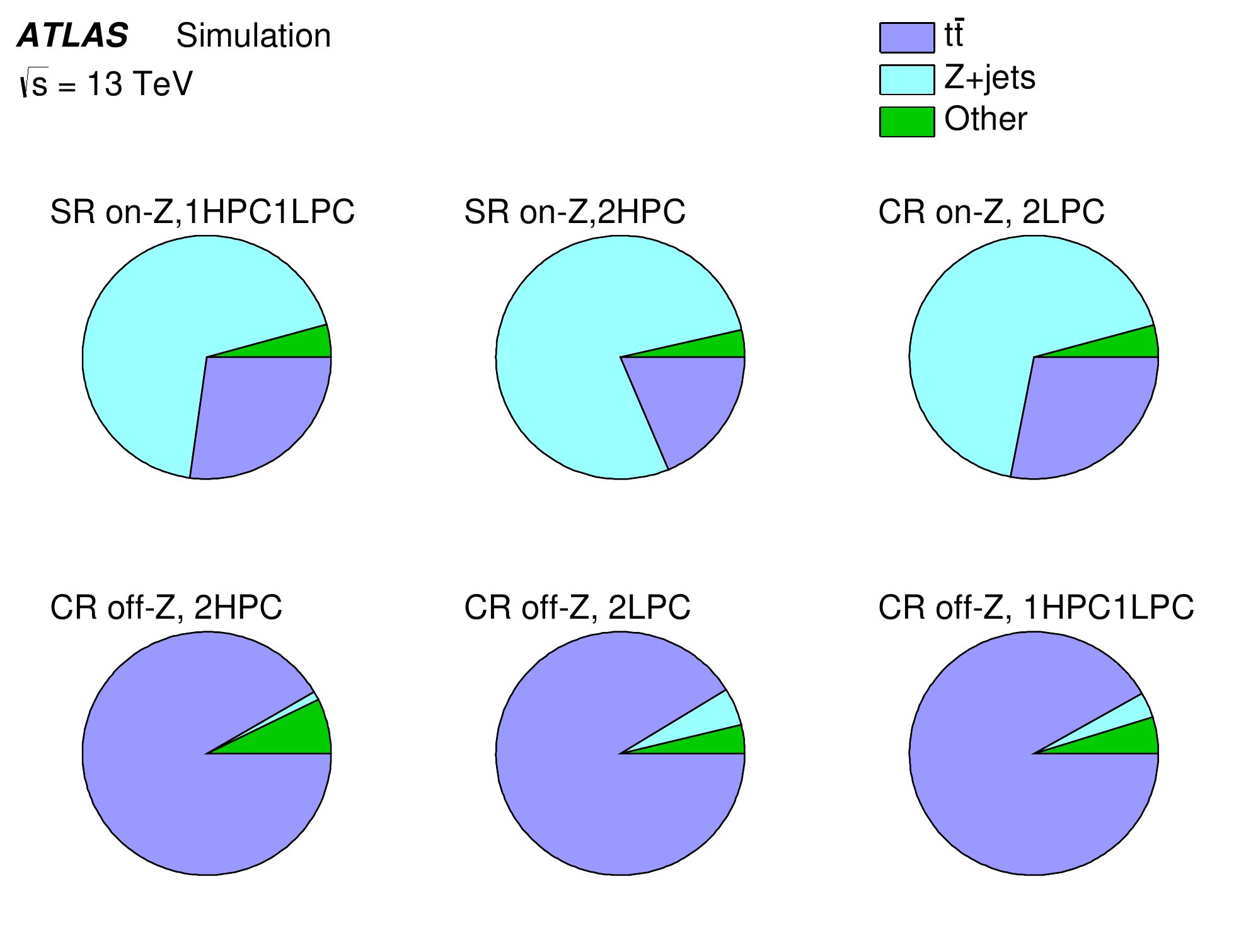}\\
\caption{Expected composition of events in each signal region (SR) and control region (CR) defined for the search. CRs have a negligible expected yield for the signal. Definitions of the regions are based on the dilepton mass and the purity of the two $\abb$ candidates. Regions labeled ``on-Z'' require the dilepton mass to be in the range $85\,\GeV\,< m_{\ell\ell} < 100\,\GeV$, while regions ``off-Z'' require the dilepton mass to be outside this window. For $\abb$ candidates, the HPC and LPC are defined using ranges of the identification BDT score, as described in Section~\ref{sec:atobb_id}.}
\label{fig:backfraction}
\end{center}
\end{figure*}

Two signal-enriched regions are defined for this search. One requires the two reconstructed $\abb$ candidates to be identified in the HPC, while the other requires one $\abb$ candidate identified in the HPC and one in the LPC. The two main sources of background for this search are top-quark pair and $\Zboson$ boson events produced in association with additional quarks or gluons.
In this search, the normalizations of these two backgrounds are measured in dedicated control regions which are selected to be enriched in the specific background. Three regions dominated by top-quark pair events are selected by requiring the dilepton mass to be outside the $\Zboson$ boson mass window, i.e.\ $m_{\ell\ell} \leq 85\,\GeV$ or $m_{\ell\ell} \geq 100\,\GeV$. These three control regions differ by the identification of the two $\abb$ candidates, with one requiring both to be in the HPC, a second requiring one $\abb$ candidate in the HPC and one in the LPC and, finally, the third requiring the two $\abb$ candidates to be identified in the LPC. The three control regions probe $\ttbar$ events produced in association with different numbers of heavy-flavor jets. A dedicated control region for $\Zboson$ boson events is formed by requiring the dilepton mass to be consistent with the $\Zboson$ boson mass and the two $\abb$ candidates in the LPC. Figure~\ref{fig:backfraction} shows the expected background yield fractions in each of the regions described here.
 
\section{Systematic uncertainties}
\label{sec:systematics}
 
Several sources of systematic uncertainty are considered. The identification efficiency for leptons is measured in $\Zboson$ boson data events using a tag-and-probe method~\cite{PERF-2017-01,PERF-2015-10}. Small residual disagreements between efficiencies in simulation and those measured in data are corrected as a function of the lepton $\pt$ and $\eta$. The uncertainties in these corrections are propagated through this search. Uncertainties in the lepton momentum scale and resolution are similarly considered.
 
Uncertainties associated with jets arise from their reconstruction and identification efficiencies. These are due to the uncertainty in the jet energy scale (JES), mass scale, energy resolution and the efficiency of the JVT requirement that is meant to remove jets from pileup. The JES and its uncertainty are derived by combining information from test-beam data, LHC collision data and simulation~\cite{PERF-2016-04}.
 
The identification efficiency for $\abb$ candidates in simulation is also corrected by using the SFs measured with the methods described in Section~\ref{sec:atobb_id}. The full covariance matrix for the 12 SFs is propagated after diagonalization in order to obtain uncorrelated sources of systematic uncertainty. Only $\abb$ candidates with $\text{BB}$ and $\text{BL}$ flavors have their identification efficiency corrected in simulation. Candidates with flavors other than $\text{BB}$ and $\text{BL}$ represent a subleading fraction of candidates selected in this analysis, mostly from $\text{BC}$ candidates. In this case, an uncertainty of 50\% per candidate is applied, similarly to the uncertainty used when measuring the identification efficiency.
 
Several sources of systematic uncertainty affecting the modeling of the relative normalization of the background sources in control and signal regions are considered. Since the $\Zjets$ background normalization is measured in a region with two $\abb$ candidates in the LPC, where a larger fraction of the candidates do not contain two $b$-quarks, an uncertainty of 50\% in the fraction of events which have two or more associated $b$-hadrons is applied.
This uncertainty is derived by comparing the level of agreement between data and simulation for $m_{\text{red}}<-40\,\GeV$ calculated with $m_a=20\,\GeV$. Similarly, for the top-quark pair background, three uncorrelated relative uncertainties of 50\% are assigned to events with one associated $b$-hadron, to events with two or more associated $b$-hadrons, and to events with associated $c$-hadrons. The number of associated hadrons in each event is determined following the procedure described in Ref.~\cite{Aaboud:2017rss}. These uncertainties are derived from a comparison of the $t\bar{t}+\textrm{heavy-flavor}$ production cross-sections predicted by {\powheg}+$\pythia$ and by $\sherpa$+\textsc{OpenLoops} at NLO~\cite{Aaboud:2017rss}.
 
Beyond the uncertainties associated with heavy-flavor fractions, several sources of systematic uncertainty affecting the relative normalization between control and signal regions are considered. The procedure closely follows the description in Ref.~\cite{Aaboud:2018iil}. For the $\ttbar$ background, it includes systematic uncertainties from variations of the factorization and renormalization scales, the PDF set used for simulation, $\alpha_{\mathrm{S}}$, the value of the top-quark mass, the choice of generator, the choice of parton shower and hadronization models, and the effects of initial- and final-state radiation. For the \Zjets{} backgrounds, additional relative uncertainties are based on variations of the factorization and renormalization scales and of the parameters used in matching the matrix element to the parton showers in the \textsc{Sherpa} simulation.
 
Uncertainties in secondary background sources are also considered, affecting their normalization in both the signal and control regions. A 50\% normalization uncertainty in the diboson background is assumed~\cite{ATL-PHYS-PUB-2016-002}. The uncertainties in the $\ttbar W$ and $\ttbar Z$ NLO cross-section predictions are 13\% and 12\%, respectively~\cite{deFlorian:2016spz,Campbell:2012dh}, and are treated as uncorrelated between the two processes. An additional modeling uncertainty for $\ttbar W$ and $\ttbar Z$, related to the choice of event generator, parton shower and hadronization models, is derived from comparisons of the nominal samples with alternative ones generated with $\sherpa$.
 
Several sources of systematic uncertainty affect the theoretical modeling of the signal. Uncertainties originate from the choice of PDFs, the factorization and renormalization scales, and the parton shower, hadronization and underlying-event models. The combined uncertainty in the expected signal yield from these sources is approximately 8\%. Higher-order corrections to the decay of the $a$-boson are small compared to the Higgs boson production uncertainties and, therefore, no additional uncertainty is included.
 
\section{Results}
\label{sec:result}
 
\begin{figure*}[htbp]
\begin{center}
\includegraphics[width=0.7\textwidth]{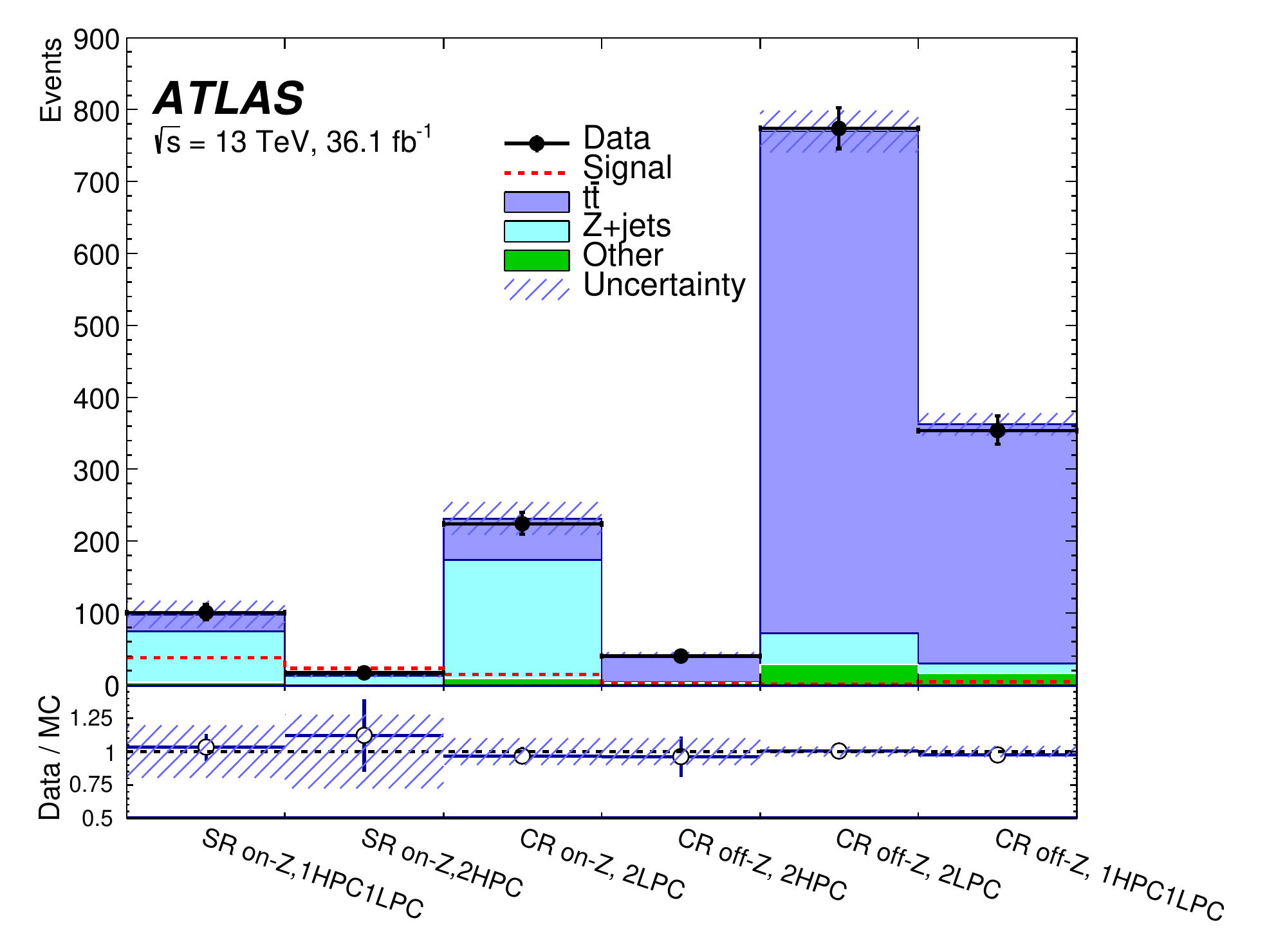}\\
\caption{Expected yields for the different background components in each signal region (SR) and control region (CR) after the profile likelihood fit to data. The expected yield for signal with $m_a = 20\,\GeV$ is calculated before the profile likelihood fit and normalized to the observed limit in the cross-section times the branching ratio for $ZH$, $H\to aa\to (\bbbar)(\bbbar)$. The data observed in each region is included for comparison. The hashed area represents the total uncertainty in the background. }
\label{fig:yields}
\end{center}
\end{figure*}

The results are obtained from a binned maximum-likelihood fit to the data using the two signal regions and four control regions.  The likelihood function is constructed from the product of Poisson probabilities in each region. The parameter of interest (POI) scales the signal $H\to aa\to (\bbbar)(\bbbar)$ yield. The overall normalizations of the $\Zjets$ and $\ttbar$ backgrounds are modeled as unconstrained nuisance parameters. Simulation is used to determine the relative yields of $\Zjets$ and $\ttbar$ backgrounds in each signal and control region. Systematic uncertainties described in Section~\ref{sec:systematics} are incorporated as nuisance parameters with Gaussian priors with a standard deviation equal to the value of the uncertainty, and these nuisance parameters multiply the product of Poisson probabilities. Uncertainties arising from the finite number of simulated events are modeled using gamma distribution priors~\cite{Barlow:1993dm}. Gamma distributions are used as a generalization of the Poisson distribution since the expected yield predicted in simulated event samples may not be an integer number. Figure~\ref{fig:yields} and Table~\ref{tbl:yields} show a comparison of data and simulation when the nuisance parameters have the values that maximize the likelihood function and only the SM background processes are considered, i.e. the POI is fixed at zero. The data in all regions agrees with the prediction within 1 standard deviation.
 
\begin{table*}[htbp]
\begin{center}
\caption{Expected yields and total uncertainty for the different background components in each signal and control region after the profile likelihood fit to data. The expected yield for signal with $m_a = 20\,\GeV$ is calculated before the profile likelihood fit and normalized to the total $ZH$ cross-section ($\BR(H\to aa\to (\bbbar)(\bbbar))=1$). The data observed in each region is included for comparison.}
\label{tbl:yields}
\vspace{0.2cm}
\small{
\begin{tabular}{l|cc|cccc}\hline\hline
&\multicolumn{2}{c}{Signal Regions}&\multicolumn{4}{|c}{Control Regions}\\
& \small{on-Z, 1HPC1LPC} & \small{on-Z, 2HPC} & \small{on-Z, 2LPC} & \small{off-Z, 2HPC} & \small{off-Z, 2LPC} & \small{off-Z, 1HPC1LPC}\\\hline
$\ttbar$ & $23.5 \pm 4.5$~~ & $2.5 \pm 0.8$ & $57.8\pm 6.9$ & $38.3\pm 4.0$~~ & $698\pm 21$~~ & $332\pm 14$\\
$\Zjets$ & $71\pm 19$ & $12.2\pm 4.1$~~ & $164 \pm 22$ & $0.5\pm 0.6$ & $44\pm 19$ & $14.0\pm 6.0$\\
Others & $3.5\pm 0.6$ & $0.4\pm 0.2$ & ~~$9.2\pm 1.1$ & $2.8\pm 0.8$ & $28.3\pm 2.4$~~ & $16.2\pm 1.8$\\\hline
Total & $98\pm 19$ & $15.2\pm 4.2$~~ & $231\pm 23$ & $41.6\pm 4.2$~~ & $770\pm 29$~~ & $362\pm 15$\\\hline\hline
Data & $101$ & $17$ & ~~$224$ & $40$ & $774$ & ~~$354$\\\hline\hline
Signal & $47 \pm 27$ & $28 \pm 11$ & ~~$18 \pm 18$ & $3.2 \pm 1.2$ & $2.1 \pm 2.1$ & ~~$5.2 \pm 3.0$ \\\hline\hline
\end{tabular}
}
\end{center}
\end{table*}
 
\begin{table*}[htbp]
\begin{center}
\caption{Impact of groups of systematic uncertainties on the expected upper limits for $m_a = 20\,\GeV$. For comparison, the statistical uncertainty impact, defined as the $1\sigma$ uncertainty of the expected upper limit after removing all nuisance parameters from the profile likelihood, is also shown.
}
\label{tbl:impacts}
\vspace{0.2cm}
\small{
\begin{tabular}{l|c}\hline\hline
Source & Impact on expected upper limit \\\hline\hline
\multicolumn{2}{c}{Systematic uncertainties}\\\hline\hline
\multicolumn{2}{c}{Objects}\\\hline
$\abb$ reconstruction and identification efficiency & ~$18\%$\\
$\abb$ energy scale and resolution & ~$13\%$\\
Lepton reconstruction and identification efficiency & $1.3\%$\\
Lepton energy scale and resolution & $0.5\%$\\\hline
\multicolumn{2}{c}{Other experimental}\\\hline
Pileup & $6.5\%$\\
Luminosity & $2.5\%$\\\hline
\multicolumn{2}{c}{Background}\\\hline
$\ttbar$ normalization & $2.0\%$\\
$\ttbar$ modeling & $7.6\%$\\
$\Zjets$ normalization & ~$13\%$\\
$\Zjets$ modeling & ~$11\%$\\
Other backgrounds & $0.8\%$\\\hline
\multicolumn{2}{c}{Signal}\\\hline
Production modeling & $3.2\%$\\\hline\hline
Statistical uncertainty & ~$43\%$\\\hline\hline
\end{tabular}
}
\end{center}
\end{table*}

Limits on the production cross-section of $ZH$, $H\to aa\to (\bbbar)(\bbbar)$ events are calculated using the test statistic $t_\mu = -2\ln({\cal L}(\mu,\hat{\hat{\theta}}_\mu)/{\cal L}(\hat{\mu},\hat{\theta}))$, where ${\cal L}$ is the likelihood described above, $\mu$ is the single POI and $\theta$ is the vector of nuisance parameters (NPs). In addition, $\hat{\mu}$ and $\hat{\theta}$ are the values which maximize the likelihood function, and $\hat{\hat{\theta}}_\mu$ are the values of the NPs which maximize the likelihood function for a given value of $\mu$~\cite{Aaboud:2018iil}.  Upper limits at 95\% C.L.\ on the production cross-section as a function of the mass hypothesis are determined using the asymptotic distribution for $t_{\mu}$~\cite{Cowan:2010js,Read:2002hq,Junk:1999kv}.
 
The impact of systematic uncertainties on the upper limits is evaluated by varying the corresponding NP when building the Asimov dataset~\cite{Cowan:2010js} used to estimate the asymptotic distribution for $t_{\mu}$. The NPs are varied by the value of their uncertainties in the fit performed to obtain ${\cal L}(\hat{\mu},\hat{\theta})$. In order to partially account for the correlation between the fitted values of the NPs, the variations are performed after diagonalizing the correlation matrix obtained in the same fit. The diagonalization is performed in blocks of NPs that share a similar origin and that may have large correlations. The impact is defined as the relative variation of the expected upper limit when the modified asymptotic distribution is used. Variations in each block are summed in quadrature and the results are shown in Table~\ref{tbl:impacts}. The number of events in each of the four control regions is the main factor in determining the impact from the unconstrained nuisance parameters that model the normalization of the $\Zjets$ and $\ttbar$ backgrounds and, therefore, their values are highly correlated. Since they are individually important for the modeling of the background yields, their impacts are reported separately. A correlation of $44\%$ is observed between the two unconstrained nuisance parameters. The impact of the statistical uncertainty is defined as the $1\sigma$ uncertainty in the expected upper limit after removing all nuisance parameters, both constrained and unconstrained, from the profile likelihood.
 
\begin{figure*}[htbp]
\begin{center}
\includegraphics[width=0.85\textwidth]{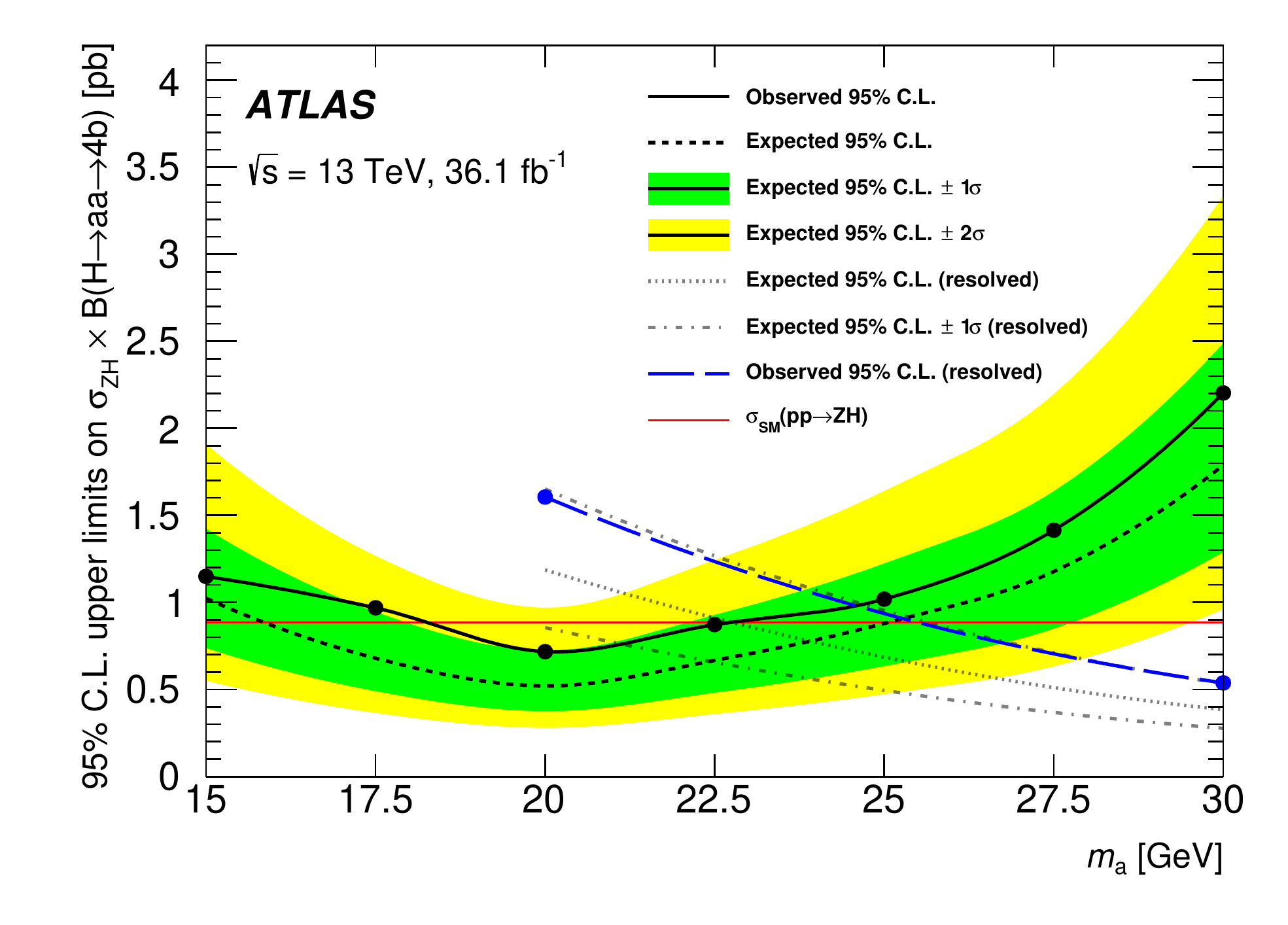}\\
\caption{Summary of the 95\% C.L.\ upper limits on $\sigma_{ZH} \BR(H
\rightarrow aa \rightarrow (\bbbar)(\bbbar))$ from this result
and a previous search targeting the higher masses, labeled resolved~\cite{Aaboud:2018iil}. The
observed limits are shown together with the expected limits. In the case of the expected limits, one- and two-standard-deviation
uncertainty bands are also displayed. The SM NNLO cross-section for $pp \rightarrow ZH$ of $0.88$~pb~\cite{deFlorian:2016spz}
is also shown.}
\label{fig:limits_resultcomparison_observedandexpected}
\end{center}
\end{figure*}

Figure~\ref{fig:limits_resultcomparison_observedandexpected} shows the exclusion limits for the production cross-section times the branching ratio for $ZH$, $H\to aa\to (\bbbar)(\bbbar)$ as a function of the $a$-boson mass hypothesis.
For comparison, the SM NNLO cross-section for $pp \rightarrow ZH$ is $\sigma_{\mathrm{SM}}(pp\rightarrow ZH)=0.88$~pb~\cite{deFlorian:2016spz}.
The figure also includes the expected exclusion limit calculated from an Asimov dataset when all the constrained nuisance parameters are fixed to their expected values and the unconstrained nuisance parameters that scale the $\Zjets$ and $\ttbar$ backgrounds are fixed to one. For $m_a=20\,\GeV$, an upper limit of $0.71\,\text{pb}$ ($0.52^{+0.31}_{-0.14}\,\text{pb}$) is observed (expected) at 95\%~C.L. The reduced sensitivity for heavier $a$-boson mass hypotheses is due to a lower acceptance caused by the increased separation of the $b$-jets, while the reduced sensitivity for lighter $a$-boson mass hypotheses is due to a lower efficiency to identify the two $b$-jets inside an $\abb$ candidate. The figure includes the results from a previous analysis targeting the higher range of $m_a$~\cite{Aaboud:2018iil}.
 
\FloatBarrier
 
\section{Conclusion}
\label{sec:conclusion}
 
A search for Higgs bosons decaying into a pair of new spin-0 particles that subsequently decay into a final state with four $b$-quarks is presented. The search uses 36 fb$^{-1}$ of $13\,\TeV$ proton--proton collision data collected by the ATLAS detector at the LHC. A dedicated strategy for reconstruction and identification of $\abb$ candidates in the mass range $15\,\GeV\,\leq m_{a} \leq 30\,\GeV$ is introduced. The measurement of the acceptance and efficiency of this strategy is described in detail and used to compare data with simulated events in regions with two $\abb$ candidates consistent with the cascade decay $H\to aa\to (\bbbar)(\bbbar)$. The dominant background sources are measured in control regions defined by relaxing some of the identification criteria. No excess of data events consistent with $H\to aa\to (\bbbar)(\bbbar)$ is observed, and upper limits at 95\% C.L.\ on the production cross-section $\sigma_{ZH} \BR(H \rightarrow aa \rightarrow (\bbbar)(\bbbar))$ are obtained as a function of the $a$-boson mass hypothesis. This novel search improves the expected limit on $\sigma_{ZH} \BR(H \rightarrow aa \rightarrow (\bbbar)(\bbbar))$ for a mass hypothesis of $m_a = 20\,\GeV$ by a factor of 2.5 when compared with the previous ATLAS result which uses the same integrated luminosity.
 
\section*{Acknowledgments}
 

We thank CERN for the very successful operation of the LHC, as well as the
support staff from our institutions without whom ATLAS could not be
operated efficiently.
 
We acknowledge the support of ANPCyT, Argentina; YerPhI, Armenia; ARC, Australia; BMWFW and FWF, Austria; ANAS, Azerbaijan; SSTC, Belarus; CNPq and FAPESP, Brazil; NSERC, NRC and CFI, Canada; CERN; ANID, Chile; CAS, MOST and NSFC, China; COLCIENCIAS, Colombia; MSMT CR, MPO CR and VSC CR, Czech Republic; DNRF and DNSRC, Denmark; IN2P3-CNRS and CEA-DRF/IRFU, France; SRNSFG, Georgia; BMBF, HGF and MPG, Germany; GSRT, Greece; RGC and Hong Kong SAR, China; ISF and Benoziyo Center, Israel; INFN, Italy; MEXT and JSPS, Japan; CNRST, Morocco; NWO, Netherlands; RCN, Norway; MNiSW and NCN, Poland; FCT, Portugal; MNE/IFA, Romania; MES of Russia and NRC KI, Russia Federation; JINR; MESTD, Serbia; MSSR, Slovakia; ARRS and MIZ\v{S}, Slovenia; DST/NRF, South Africa; MICINN, Spain; SRC and Wallenberg Foundation, Sweden; SERI, SNSF and Cantons of Bern and Geneva, Switzerland; MOST, Taiwan; TAEK, Turkey; STFC, United Kingdom; DOE and NSF, United States of America. In addition, individual groups and members have received support from BCKDF, CANARIE, Compute Canada and CRC, Canada; ERC, ERDF, Horizon 2020, Marie Sk{\l}odowska-Curie Actions and COST, European Union; Investissements d'Avenir Labex, Investissements d'Avenir Idex and ANR, France; DFG and AvH Foundation, Germany; Herakleitos, Thales and Aristeia programmes co-financed by EU-ESF and the Greek NSRF, Greece; BSF-NSF and GIF, Israel; La Caixa Banking Foundation, CERCA Programme Generalitat de Catalunya and PROMETEO and GenT Programmes Generalitat Valenciana, Spain; G\"{o}ran Gustafssons Stiftelse, Sweden; The Royal Society and Leverhulme Trust, United Kingdom.
 
The crucial computing support from all WLCG partners is acknowledged gratefully, in particular from CERN, the ATLAS Tier-1 facilities at TRIUMF (Canada), NDGF (Denmark, Norway, Sweden), CC-IN2P3 (France), KIT/GridKA (Germany), INFN-CNAF (Italy), NL-T1 (Netherlands), PIC (Spain), ASGC (Taiwan), RAL (UK) and BNL (USA), the Tier-2 facilities worldwide and large non-WLCG resource providers. Major contributors of computing resources are listed in Ref.~\cite{ATL-SOFT-PUB-2020-001}.
 
 
\appendix
\printbibliography
 
\clearpage 
 
\begin{flushleft}
\hypersetup{urlcolor=black}
{\Large The ATLAS Collaboration}

\bigskip

\AtlasOrcid[0000-0002-6665-4934]{G.~Aad}$^\textrm{\scriptsize 102}$,    
\AtlasOrcid[0000-0002-5888-2734]{B.~Abbott}$^\textrm{\scriptsize 128}$,    
\AtlasOrcid{D.C.~Abbott}$^\textrm{\scriptsize 103}$,    
\AtlasOrcid[0000-0002-2788-3822]{A.~Abed~Abud}$^\textrm{\scriptsize 36}$,    
\AtlasOrcid[0000-0002-1002-1652]{K.~Abeling}$^\textrm{\scriptsize 53}$,    
\AtlasOrcid[0000-0002-2987-4006]{D.K.~Abhayasinghe}$^\textrm{\scriptsize 94}$,    
\AtlasOrcid[0000-0002-8496-9294]{S.H.~Abidi}$^\textrm{\scriptsize 166}$,    
\AtlasOrcid[0000-0002-8279-9324]{O.S.~AbouZeid}$^\textrm{\scriptsize 40}$,    
\AtlasOrcid{N.L.~Abraham}$^\textrm{\scriptsize 155}$,    
\AtlasOrcid[0000-0001-5329-6640]{H.~Abramowicz}$^\textrm{\scriptsize 160}$
% ...
\bigskip
\\

% ...
$^{36}$CERN, Geneva; Switzerland.\\
% ...
$^{40}$Niels Bohr Institute, University of Copenhagen, Copenhagen; Denmark.\\
% ...
$^{53}$II. Physikalisches Institut, Georg-August-Universit\"{a}t G\"ottingen, G\"ottingen; Germany.\\
% ...
$^{94}$Department of Physics, Royal Holloway University of London, Egham; United Kingdom.\\
% ...
$^{102}$CPPM, Aix-Marseille Universit\'e, CNRS/IN2P3, Marseille; France.\\
$^{103}$Department of Physics, University of Massachusetts, Amherst MA; United States of America.\\
% ...
$^{128}$Homer L. Dodge Department of Physics and Astronomy, University of Oklahoma, Norman OK; United States of America.\\
% ...
$^{155}$Department of Physics and Astronomy, University of Sussex, Brighton; United Kingdom.\\
% ...
$^{160}$Raymond and Beverly Sackler School of Physics and Astronomy, Tel Aviv University, Tel Aviv; Israel.\\
% ...
$^{166}$Department of Physics, University of Toronto, Toronto ON; Canada.\\
% ...

\end{flushleft}


\end{document}